\documentclass[prb,twocolumn,showpacs]{revtex4}
\usepackage{graphicx}
\usepackage{amsmath,amssymb,amsfonts}
\usepackage{hyperref}
\usepackage{bm}

\newcommand{\bQ}{\bm Q}
\newcommand{\tJ}{\tilde{J}}

\newcommand{\Ham}{\mathcal{H}}

\begin{document}

\title{Orbital ordering and unfrustrated $(\pi,0)$ magnetism from degenerate double exchange in the iron pnictides}
\author{Weicheng Lv, Frank Kr\"uger, and Philip Phillips}
\affiliation{Department of Physics, University of Illinois, 1110 West Green Street, Urbana, Illinois 61801, USA}
\date{\today}

\begin{abstract}
The magnetic excitations of the iron pnictides are explained within a degenerate double-exchange model. The local-moment spins are coupled by superexchanges
$J_1$ and $J_2$ between nearest and next-nearest neighbors, respectively, and interact with the itinerant electrons of the degenerate $d_{xz}$ and $d_{yz}$ orbitals
via a ferromagnetic Hund exchange. The latter stabilizes $(\pi,0)$ stripe antiferromagnetism due to emergent ferro-orbital order and the resulting kinetic energy gain
by hopping preferably along the ferromagnetic spin direction. Taking the quantum nature of the spins into account, we calculate the magnetic excitation spectra
in the presence of both, super- and double-exchange.
A dramatic increase of the spin-wave energies at the competing N\'eel ordering wave vector is found, in agreement with recent neutron scattering data.
The spectra are fitted to a spin-only model with a strong spatial anisotropy and additional longer ranged couplings along the ferromagnetic chains. Over a realistic
parameter range, the effective couplings along the chains are negative
corresponding to unfrustrated stripe antiferromagnetism.
\end{abstract}


\maketitle

\section{Introduction}
The discovery of superconductivity in the
pnictides\cite{Kamihara+08,Chen+08a,Chen+08b,Ren+08,Wen+08,Rotter+08}
with transition temperatures challenging those of single-layer, high-$T_c$ cuprates, immediately raised the question of whether, despite all their differences,
 the two classes of materials share the same key mechanism for superconductivity.\cite{Kivelson+08,Zaanen09,Tesanovic09} Arguably the most striking similarity
 is that superconductivity emerges upon doping antiferromagnetically ordered parent compounds. In the pnictides, however, the magnetic ordering is unusual,
 characterized by an antiferromagnetic arrangement of ferromagnetic chains, corresponding to an in-plane ordering wave vector
$\bQ=(\pi,0)$.\cite{Cruz+08,Zhao+08b,Zhao+08c,Kimber+08,Goldman+08,Zhao+08d,Huang+08,Wilson+09,Chen+08c,Dong+08}
Whereas the  pnictides are metallic, the parent cuprates are Mott insulators. The conductivities in the pnictides are typical of bad metals, suggesting
that electronic  correlations\cite{Wu+08, Haule+08,Craco+08,Nakamura+08,Vildosola+08,Anisimov+08,Yang+09,Si+08,Fang+08,Xu+08b} are
crucial.  By contrast, the local-density approximation  seems to be adequate\cite{Mazin+08} to describe their band structure.

As a result of this dichotomy, both itinerant-magne\-tism\cite{Mazin+08,Kuroki+08,Raghu+08,Chubukov+08,Cvetkovic+09,Wang+09,Brydon+09,Knolle+10,Zhang+10} and
local-moment\cite{Yildirim+08,Si+08,Fang+08,Xu+08b,Kruger+09,Jiang+09,Uhrig+09,Applegate+10,Schmidt+10} scenarios have been suggested to explain
the unusual stripe antiferromagnetism. Although the former weak-coupling theories which attribute the magnetism to a spin-density-wave instability of a nested Fermi
surface, can explain the magnetic low-energy excitations around the correct ordering wave vector, they fail to describe the spectra at higher energies which
have been measured in great detail by inelastic neutron scattering.\cite{Zhao+08,Zhao+09} In particular, the itinerant scenarios can not explain the observed maximum
of the dispersion at $(\pi,\pi)$ but rather suggest that the excitations rapidly dissolve into a particle-hole continuum\cite{Knolle+10} which has not been found
experimentally up to energies of 200meV. So far, a consistent description of the excitations over the entire energy range has been obtained only  by using suitably parametrized Heisenberg models.

Because of the positions of the arsenic ions above or below the iron plaquettes, such a spin-only model is expected to be strongly frustrated with comparable,
antiferromagnetic superexchanges $J_1$ and $J_2$ between nearest and next-nearest neighbors. In this regime, the model indeed exhibits long-range
stripe-antiferromagnetic order,\cite{Chandra+88,Henley+89,Chandra+90} and the strong frustration and proximity to a continuous magnetic phase transition might explain
why the observed magnetic moments are relatively small.\cite{Si+08,Uhrig+09} Interestingly, the neutron-scattering experiments tell a radically different story. The
spin-wave velocities indicate that the exchange coupling along the ferromagnetic spin direction is much smaller than the one perpendicular to the
chains,\cite{Zhao+08} $J_{1y}\ll J_{1x}$. More recently, it has been argued\cite{Zhao+09} that the observed maximum
of the dispersion at $(\pi,\pi)$ requires an even slightly ferromagnetic exchange $J_{1y}<0$ corresponding to an unfrustrated spin model in contrast to early
claims\cite{Si+08} of high frustration.

What might be the cause of such a strong spatial anisotropy? In fact, before the magnetic order sets in, a structural transitions occurs at which the two in-plane lattice constants
become inequivalent.  The structural and magnetic transition are clearly separated in the so-called `1111 compounds',\cite{Cruz+08,Zhao+08b,Zhao+08c,Kimber+08}  whereas they occur at the same temperature in the `122 family'.\cite{Goldman+08,Zhao+08d,Huang+08,Wilson+09} However, inspecting the numbers, it appears that the orthorhombic lattice distortion is too small, by two orders of magnitude, to explain the magnetic anisotropy.\cite{Schmidt+10}

To this end, some have proposed\cite{Kruger+09,Lv+09,Lee+09,Chen+09} that orbital-ordering physics of a similar kind as in the manganite transition-metal oxides not
only provides a mechanism for the lattice distortion but more importantly explains the strong in-plane anisotropies. In particular, it has been argued\cite{Kruger+09} that due
to an orbital degeneracy, the localized limit is described by a complicated spin-orbital superexchange (Kugel-Khomskii) model rather than by a Heisenberg Hamiltonian.
Further, it was shown that the stripe antiferromagnetism is stabilized by  ferro-orbital order which breaks the in-plane lattice symmetry and induces a strong
anisotropy between the magnetic exchange couplings.

Since the $C_4$ lattice symmetry is broken by the orbital order and the accompanying orthorhombic distortion, the electronic structure must reflect this spatial anisotropy
with reduced symmetry.\cite{Lee+09b} Indeed, such an anisotropic electronic state has been confirmed recently in scanning-tunneling-microscopy (STM)\cite{Chuang+10}
and in-plane resistivity measurements.\cite{Tanatar+10,Chu+10}  Dramatic Fermi-surface reconstructions at the structural transition,\cite{Shimojima+10} as well as enormous
transport\cite{McGuire+08} and phonon\cite{Akrap+09} anomalies, have been interpreted as indirect evidence for orbital ordering.  Hence, on this interpretation, it is the orbital ordering that underlies the electronic anisotropy, not an inherent anisotropy due entirely to the electrons, indicative of a true nematic state.

Consequently, an open problem with the pnictides is the role itineracy and local physics play in mediating the apparently unfrustrated anisotropic magnetism.
In this work, we start from the idea of the `local-itinerant dichotomy' \cite{Wu+08,Kou+09,Dai+09} of the iron pnictides and motivate an effective degenerate double-exchange
(DDEX) model, very similar to the ones used to describe metallic manganites with orbital degeneracies.\cite{Brink+99a,Brink+99b,Brink+01} To be more precise, we assume
a ferromagnetic Hund coupling between the itinerant bands of the doubly-degenerate $d_{xz}$ and $d_{yz}$ orbitals and the local moments, which are described by the aforementioned $J_1$-$J_2$ Heisenberg model. 
In the context of the manganites it has been shown\cite{Brink+99a,Brink+99b,Brink+01} that such DDEX models exhibit
phases with long-range stripe-antiferromagnetic order. Despite the antiferromagnetic couplings between local-moment spins, ferromagnetic spin chains are 
stabilized by emergent ferro-orbital order. In this phase, the itinerant electrons are directed predominantly along the chains which minimizes the kinetic energy and gives rise to
a highly anisotropic electronic state. Moreover, the double exchange is expected to strongly suppress the effective coupling between local moments
along the chains, and possibly to make it even ferromagnetic.\cite{Senff+06}

In this work, we analyze the effective DDEX model and indeed find that the orbitally ordered $(\pi,0)$ antiferromagnet is stable over a realistic parameter range for the 
parent iron-pnictide materials. In particular, the seizable next-nearest neighbor
superexchange $J_2$ further stabilizes this phase. Whereas these results are to a large extent not surprising given the similarities with the manganites, the magnetic excitation
spectra so far have been calculated only for a ladder system.\cite{Neuber+06} For the manganites, the DDEX model is usually simplified\cite{Brink+99a,Brink+99b,Brink+01}
by treating the core spins as classical and by taking the limit of large or infinite Hund's coupling $J_\mathrm{H}$ which is not justified for the pnictides. Here, we instead focus on the
regime of small and intermediate $J_\mathrm{H}$ and develop the tools to calculate the magnetic excitation spectra in the presence of both, super- and double-exchange to
linear-spin-wave order. The spectra are found to be in good agreement with the neutron scattering data.\cite{Zhao+09} In particular, in some parameter
space the double exchange along the ferromagnetic chains can overcompensate the bare antiferromagnetic superexchange
as suggested by the experiment.

This paper is organized as follows. In Sec.~\ref{sec:model}, we
construct the local-itinerant, DDEX model. Sec.~\ref{sec:method} deals with the methods
we use to calculate the magnetic excitation spectra. 
In Sec.~\ref{sec:results}, we summarize our results,
including ferro-orbital ordering, the spin-wave dispersions, and the magnetic anisotropies. Finally Sec.~\ref{sec:disc} discusses several aspects of our theory
and validates it's applicability to the pnictides.

\section{Model}
\label{sec:model}

In this Section, we proceed to motivate the DDEX model for the pnictides. This model accounts for the presence of local moments, as suggested by
the neutron-scattering experiments, as well as itinerant electrons responsible for the bad-metal behavior of the parent compounds. Moreover, the orbital degeneracy in
combination with Hund's coupling between electronic and spin degrees of freedom gives rise to orbital-ordering physics beyond simple band-structure theory. The
Hamiltonian consists of three parts,
\begin{equation}
\Ham  = \Ham_\mathrm{loc} + \Ham_\mathrm{it} + \Ham_\mathrm{H},
\label{eq:start}
\end{equation}
where $\Ham_\mathrm{loc}$ describes the superexchange couplings between local moments, $\Ham_\mathrm{it}$ the itinerant electrons of the degenerate $d_{xz}$ and $d_{yz}$
orbitals, and $\Ham_\mathrm{H}$ the ferromagnetic Hund coupling between local moments and  itinerant electrons. In order for this model to be valid, Hund's coupling $J_\mathrm{H}$ should be small compared to the tetrahedral crystal-field splitting between the $t_{2g}$ and $e_g$ multiplets, but larger than the tetragonal
splitting between the $d_{xy}$ orbital and the degenerate $d_{xz}$,$d_{yz}$ doublet.\cite{Kruger+09}

\begin{figure}[t]
  \centering
  \includegraphics[width=8cm]{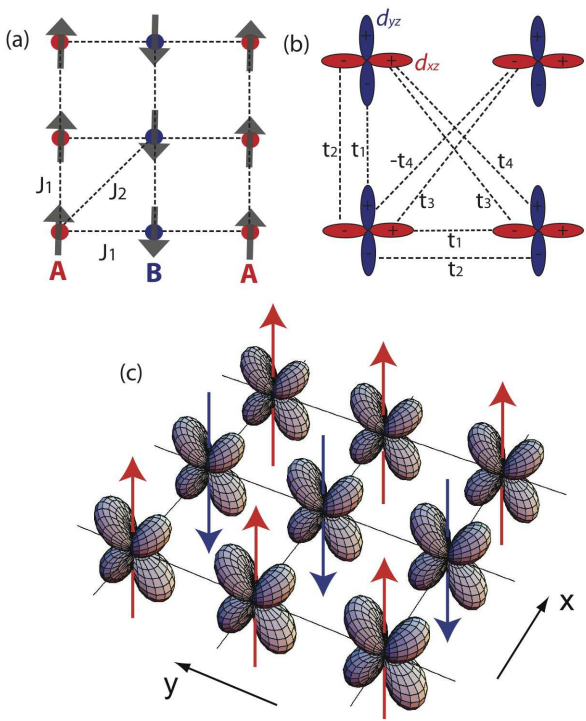}
  \caption{(Color online) Illustration of the degenerate double-exchange model for the pnictides.  (a) The local moments are coupled by nearest- and next-nearest-neighbor
  exchanges $J_1$ and $J_2$, respectively, and interact via a ferromagnetic Hund coupling $J_\mathrm{H}$ with the itinerant electrons of the degenerate $d_{xz}$, $d_{yz}$
  orbitals. (b) Illustration of the hopping parameters in a two-band model of these orbitals (shown as projections in the plane).  (c) Resulting ferro-orbital order
  which stabilizes $(\pi,0)$ antiferromagnetism by directing the kinetic energy of
  the electrons along the ferromagnetic spin direction.}
  \label{fig:schematic}
\end{figure}

The local moments with spin $S$ are coupled by superexchanges  $J_1$ and $J_2$ between nearest and next-nearest neighbors, respectively. The corresponding Heisenberg
Hamiltonian reads
\begin{equation}
\Ham_\mathrm{loc}  =  \frac{J_1}{S^2} \sum_{\langle i,j \rangle} \bm{S}_i \cdot \bm{S}_j + \frac{J_2}{S^2} \sum_{\langle \langle i,j \rangle \rangle} \bm{S}_i \cdot \bm{S}_j,
\label{eq:heisenberg}
\end{equation}
where, for convenience, the superexchanges are measured in units of $S^2$. Likewise,  Hund's exchange $J_\mathrm{H}$ which couples the electron spins to the
local moments will be measured in units of $S$. This convention will facilitate our large-$S$ expansion later. The superexchanges are mediated by virtual hopping processes
via the $p$-orbitals of the arsenic ions which have alternating positions above and below the iron plaquettes. Certainly, a quantitative comparison of the exchange couplings
would require knowledge of the two different Fe-As-Fe bond angles and the precise shape of the orbitals. Assuming that the virtual processes for the
nearest-neighbor and diagonal bonds involve roughly the same energies, we estimate $J_1\approx 2 J_2$ since two exchange paths via different arsenic ions contribute
to $J_1$. Therefore, we expect the Heisenberg model (\ref{eq:heisenberg}) to be strongly frustrated and potentially in the quantum disordered regime.

The itinerant electrons of the degenerate $d_{xz}$ and $d_{yz}$ orbitals are described by a tight-binding Hamiltonian
\begin{equation}
\Ham_\mathrm{it}  =   -\sum_{ij,\alpha\beta,\nu} t_{ij}^{\alpha\beta} c_{i\alpha \nu}^\dagger c_{j \beta \nu},
\end{equation}
where $c_{i\alpha \nu}^\dagger$ creates an electron with spin $\nu$ at site $i$ on orbital $\alpha$. The hopping integrals $t_{ij}^{\alpha\beta}$ are illustrated in
Fig.~\ref{fig:schematic}(b) and defined in the same way as in Ref.~\onlinecite{Raghu+08}. For simplicity, we neglect inter- and intra-orbital Coulomb repulsions\cite{Oles+83}
between the itinerant electrons. This is justified since on the level of the effective DDEX model, the local moments are formed as a consequence of strong correlations
whereas the residual charge carriers should be viewed as weakly interacting quasiparticles.
We will assume $t_1$ to be the dominant hopping because a larger wave-function overlap is expected when the orbitals point towards one another.
However, the precise shape of the orbitals is not determined by geometry but depends on quantum chemistry, for example, on the amount of hybridization between the
Fe $d$- and As $p$-orbitals.\cite{Wu+08} Here we simply denote the orbitals by $d_{xz}$ and $d_{yz}$ because of the spatial symmetry shared with the
atomic Fe orbitals. Recently, it has been suggested\cite{Lee+09} that the hybridization leads to a strong deformation of the orbitals which make the $\pi$ overlap,
$t_2$ the dominant one. We point out that our results persist for the exchange of $t_1$ and $t_2$, the only difference being
that the orbital polarization will be inverted in order to maximize the overlap along the ferromagnetic spin direction.

We do not attempt to fit our hopping parameters to reproduce the
electron and hole pockets as has been done in previous two-orbital models\cite{Raghu+08,Moreo+09} at
the level of a tight-binding approximation. Such parameters are inherently arbitrary since the hopping amplitudes are not uniquely determined by a particular constant
energy cut\cite{Moreo+09} and because Coulomb interactions and Hund's exchange are of the order of the electronic
bandwidth.\cite{Haule+08,Craco+08,Nakamura+08,Vildosola+08,Anisimov+08,Yang+09} Moreover, the Fermi surfaces in the antiferromagnetically ordered phase have not
been clearly established yet.

Finally the local moments and the itinerant electrons interact by a ferromagnetic Hund coupling,
\begin{equation}
\Ham_\mathrm{H} = - \frac{J_\mathrm{H}}{2S} \sum_{i,\alpha, \nu\nu^\prime} \bm{S}_i \cdot c_{i \alpha \nu}^\dagger \bm{\sigma}_{\nu \nu^\prime} c_{i \alpha \nu^\prime},
\label{eq:hund}
\end{equation}
where $\bm{\sigma}_{\nu \nu^\prime}=(\sigma^x,\sigma^y,\sigma^z)_{\nu \nu^\prime}$ with $\sigma^\alpha$ the standard Pauli matrices
and $J_\mathrm{H} > 0$. As mentioned before, Hund's exchange is measured in units of the local moment $S$. We note that similar models have been proposed in
the context of the pnictides.\cite{Kou+09,Dai+09} However, the orbital degeneracy which is the pre-requisite for orbital-ordering physics and the resulting spatial anisotropies
has not been included.

\section{Method}
\label{sec:method}

In this Section, we outline the approximations and transformation we employ to analyze the complicated DDEX model for the pnictides.
In similar models for the maganites, the problem is typically simplified by treating the local moments as classical spins and assuming an infinitely strong Hund
exchange.\cite{Brink+99a,Brink+99b,Brink+01} In the pnictides, these
approximations are not justified since the Hund coupling is of the order of the electronic
bandwidth and since the local moments are small and presumably best described by the extreme quantum limit, $S=1/2$. Moreover, our goal is the calculation of
the magnetic excitation spectra  which will require the inclusion of quantum fluctuations of the spins.
Although the spins are small and the Heisenberg model $\Ham_\mathrm{loc}$ is strongly frustrated, it is reasonable to treat the local moments on the
level of linear spin-wave theory since the double exchange is expected to lead to a dramatic stabilization of the magnetic order. Moreover, it has been argued that the $1/S$
expansion is much better behaved for $(\pi,0)$ order as compared to conventional $(\pi,\pi)$ N\'eel antiferromagnetism.\cite{Uhrig+09}

Since Hund's coupling $\Ham_\mathrm{H}$ does not conserve the magnons describing the spin-wave excitation of the isolated local moments, we perform
a canonical transformation in order to identify the true magnons of the coupled system. Readers not interested in the details of this
calculation can skip immediately to the results, Sec.~\ref{sec:results}.

\subsection{Operator rotations}

Following the standard treatment of the antiferromagnetic Heisenberg model, we perform the spin rotation $S_i^x = \tilde{S}_i^x$, $S_i^y = \kappa_i \tilde{S}_i^y$,
and $S_i^z = \kappa_i \tilde{S}_i^z$ where $\kappa_i = \exp{(i\bm{Q} \cdot \bm{r}_i)}= \pm 1$
for sublattices A and B, respectively [see Fig.~\ref{fig:schematic}(a)]. Representing the rotated spin operators $\tilde{\bm{S}}_i$ by Holstein-Primakoff (HP) bosons
$a_i$, $a_i^\dagger$, to the leading order, $\tilde{S}_i^z  = S - a_i^\dagger a_i$,  $\tilde{S}_i^+ = \sqrt{2S} a_i$, and  $\tilde{S}_i^- =  a_i^\dagger \sqrt{2S}$
($\tilde{S}_i^\pm=\tilde{S}_i^x\pm i\tilde{S}_i^y$), we immediately derive the following expression for $\Ham_\mathrm{loc}$ in the linear spin-wave approximation,
\begin{eqnarray}
\Ham_\mathrm{loc}^\mathrm{sw} & = & \sum_q  \left[A_0(q) \left(a_q^\dagger a_q+a_{-q} a_{-q}^\dagger  \right) \right.\nonumber\\
& & \left.+ B_0(q) \left( a_q^\dagger a_{-q}^\dagger +  a_{-q} a_q \right)\right],
\label{eq:hb}
\end{eqnarray}
where $A_0(q)=(J_1 \cos{q_y} + 2 J_2)/S$ and $B_0(q)=(J_1 \cos{q_x} + 2 J_2 \cos{q_x} \cos{q_y})/S$.

In order to leave Hund's coupling term $\Ham_\mathrm{H}$
invariant, we perform exactly the same rotation of the electron spins $\bm{s}_{i\alpha}=\frac 12\sum_{\nu\nu^\prime}c_{i \alpha \nu}^\dagger \bm{\sigma}_{\nu \nu^\prime} c_{i \alpha \nu^\prime}$. This is achieved by transforming the fermion operators  as $c_{i \alpha \nu}  =  \tilde{c}_{i \alpha \nu}$ for sites $i$ on sublattice A and
$c_{i \alpha \nu} =  \tilde{c}_{i \alpha \bar{\nu}}$ on sublattice B. In the latter expression, we have defined $\bar{\nu} = \downarrow$ for $\nu = \uparrow$ and vice versa.
In terms of the HP boson creation and annihilation operators $a_i$, $a_i^\dagger$ and rotated fermion operators $\tilde{c}_{i \alpha \nu}$, $\tilde{c}^\dagger_{i \alpha \nu}$ Hund's exchange can be written as
\begin{eqnarray}
\Ham_\mathrm{H} & = & \Ham_\mathrm{H}^{(0)} + \Ham_\mathrm{H}^{(1)} + \Ham_\mathrm{H}^{(2)}, \\
\Ham_\mathrm{H}^{(0)} & = & - \frac{J_\mathrm{H} }{2} \sum_{k, \alpha, \nu} \nu \tilde{c}_{k \alpha \nu}^\dagger \tilde{c}_{k \alpha \nu}, \\
\Ham_\mathrm{H}^{(1)} & = & - \frac{J_\mathrm{H}} {\sqrt{2S}}  \sum_{kq,\alpha} \left( a_q \tilde{c}_{k+q,\alpha\downarrow}^\dagger \tilde{c}_{k\alpha\uparrow} + \textrm{h.c.} \right), \\
\Ham_\mathrm{H}^{(2)} & = &  \frac{J_\mathrm{H}}{2S} \sum_{k, q q^\prime, \alpha \nu} \nu a_q^\dagger a_{q^\prime} \tilde{c}_{k-q,\alpha\nu}^\dagger
\tilde{c}_{k-q^\prime,\alpha\nu},
\end{eqnarray}
where $\nu = \pm 1$ for up and down spins, respectively. Note that $\Ham_\mathrm{H}^{(0)}$ only involves the electronic operators and represents the zeroth-order corrections to the
electron energies by the classical background stripe antiferromagnetism. $\Ham_\mathrm{H}^{(1)}$ and $\Ham_\mathrm{H}^{(2)}$ are the couplings between the electrons and the
HP bosons, linear and quadratic in the boson operators. In the following, we include the term $\Ham_\mathrm{H}^{(0)}$ in the itinerant-electron Hamiltonian, yielding the
effective free-electron Hamiltonian
\begin{eqnarray}
\Ham_\mathrm{e} & = & \Ham_\mathrm{it} + \Ham_\mathrm{H}^{(0)}\nonumber\\
& =  & \sum_{k,\alpha ,\nu} \left[\left( \varepsilon_1^{\alpha} (k)  - \nu \frac{J_\mathrm{H}}{2} \right)  \tilde{c}_{k \alpha \nu}^\dagger \tilde{c}_{k \alpha \nu}
\right.\nonumber \\
& &  \left.\phantom{\frac{J_\mathrm{H}}{2}}+\varepsilon_2^{\alpha}(k) \, \tilde{c}_{k \alpha \nu}^\dagger \tilde{c}_{k \alpha \bar{\nu}}
 + \varepsilon_3(k) \, \tilde{c}_{k \alpha \nu}^\dagger \tilde{c}_{k \bar{\alpha} \bar{\nu}}\right],
\label{eq:heq}
\end{eqnarray}
where $\bar{\alpha} = yz$ for $\alpha = xz$ and vice versa. We have defined $\varepsilon_1^{xz} (k)  =  -2t_2 \cos{k_y}$, $\varepsilon_1^{yz}(k) = -2t_1 \cos{k_y}$,
$\varepsilon_2^{xz}(k)  =  -2t_1 \cos{k_x} - 4t_3\cos{k_x}\cos{k_y}$,  $\varepsilon_2^{yz}(k)  =  -2t_2 \cos{k_x} - 4t_3\cos{k_x}\cos{k_y}$, and $\varepsilon_3(k) = -t_4\sin{k_x}\sin{k_y}$.

\subsection{Canonical transformation}

Apparently, the interaction term $\Ham_\mathrm{H}^{(1)}$ is linear in
the HP-boson operators, which shows that these bosons do not represent the Goldstone modes of the system, namely the transverse fluctuations of the total staggered magnetic moments, which consist of not only the local moments, but also the spins of the itinerant electrons. In order to correctly identify the true magnons and carry out the subsequent spin-wave calculations, we need to perform a canonical transformation  of the original Hamiltonian
$\Ham=\Ham_\mathrm{loc}^\mathrm{sw} +\Ham_\mathrm{e} +\Ham_\mathrm{H}^{(1)}+\Ham_\mathrm{H}^{(2)}$,
\begin{eqnarray}
\Ham^\prime & = & e^{\Delta} \Ham e^{-\Delta} \nonumber \\
	& = & \Ham + \left[\Delta , \Ham \right] + \frac{1}{2} \left[\Delta, \left[ \Delta, \Ham\right] \right] + \dots
\label{eq:canon}
\end{eqnarray}
with $\Delta$ a suitable anti-Hermitian operator, $\Delta^\dagger = - \Delta$, such that in the transformed $\Ham^\prime$, the terms linear in $a_i$'s are eliminated.
Similar canonical transformations\cite{Nagaev+98,Golosov+00} and equivalent perturbative methods\cite{Shannon+02} have been used to explain
ferromagnetism in double-exchange models with a single itinerant band.   Up to the leading order, the transformation is determined by
\begin{equation}
\left[\Delta, \Ham_\mathrm{e} \right] + \Ham_\mathrm{H}^{(1)} = 0.
\label{eq:delta}
\end{equation}

To find the $\Delta$ satisfying (\ref{eq:delta}), we first diagonalize $\Ham_\mathrm{e}$ by a unitary transformation
$\tilde{c}_{k \alpha \nu} = \sum_n U_{\alpha \nu}^n (k) d_{nk}$, yielding $\Ham_\mathrm{e} = \sum_{n,k}{E_n(k)d_{nk}^{\dagger}d_{nk}}$. Here, $n$ labels the four
electronic bands arising after diagonalization from the two orbital and two spin degrees of freedom.
In the new basis of $d_{nk}$, it is easy to verify that (\ref{eq:delta}) is solved by
\begin{eqnarray}
\Delta & = & \frac{J_\mathrm{H}}{\sqrt{2S}} \sum_{kq,mn,\alpha} \left(\frac{a_q d_{m,k+q}^{\dagger} d_{nk}}{E_n(k) - E_m(k+q)} \right.\nonumber \\
& &\left. \phantom{\frac{d_k^\dagger}{E_n}}\times \, U^{m\ast}_{\alpha \downarrow} (k+q) U_{\alpha \uparrow}^n (k)  -  \textrm{h.c.}\right).
\end{eqnarray}

After the canonical transformation (\ref{eq:canon}), the Hamiltonian up to order $1/S$ reads $\Ham^\prime = \Ham_\mathrm{e} + \Ham_\mathrm{loc}^\mathrm{sw} +
\Ham_\mathrm{H}^{(2)} + \Ham_2^\prime$, where $\Ham_2^\prime= [\Delta,\Ham_\mathrm{H}^{(1)}]+\frac 12[\Delta[\Delta, \Ham_\mathrm{e} ]]=
\frac 12 [\Delta,\Ham_\mathrm{H}^{(1)}]$. The commutators $[\Delta, \Ham_\mathrm{loc}^\mathrm{sw}]$ and $[\Delta, \Ham_\mathrm{K}^{(2)}]$ are of higher orders
in $1/S$ and the boson operators, and thus can be dropped in the linear spin-wave approximation. The contributions $\Ham_\mathrm{H}^{(2)}$ and
$\Ham_2^\prime$ are bilinear in both, the bosonic and fermionic operators. After  taking the expectation values of the electronic operators with respect to the
diagonal electronic Hamiltonian $\Ham_\mathrm{e}$, we obtain the final spin-wave Hamiltonian
\begin{eqnarray}
\Ham^\mathrm{sw} & = & \Ham_\mathrm{loc}^\mathrm{sw}+\left\langle \Ham_\mathrm{H}^{(2)}+\frac 12 [\Delta,\Ham_\mathrm{H}^{(1)}]\right\rangle_\mathrm{e}\nonumber\\
& = &  \sum_q  \left[A(q)\left(a_q^\dagger a_q+a_{-q} a_{-q}^\dagger  \right) \right.\nonumber\\
& & \left.+ B(q) \left( a_q^\dagger a_{-q}^\dagger +  a_{-q} a_q \right)\right]
\label{eq:swfinal}
\end{eqnarray}
with $A(q)=A_0(q)+A_1+A_2(q)$ and $B(q)=B_0(q)+B_2(q)$. The constant
`self-energy' correction, $A_1$ arises from $H_\mathrm{K}^{(2)}$ whereas $H_2^\prime=\frac 12 [\Delta,\Ham_\mathrm{H}^{(1)}]$ generates  momentum-dependent corrections to 
both, the `self-energy' and the `anomalous amplitude', which are denoted as $A_2(q)$ and $B_2(q)$.
These corrections are expressed as
\begin{eqnarray}
	A_1 & = & \frac{J_\mathrm{H}}{2S} \sum_{k,n} f_n(k) \sum_{\alpha,\nu} \nu\left\vert U_{\alpha \nu}^n (k) \right\vert ^2, \\
	A_2(q) & = & \frac{J_\mathrm{H}^2 }{2S} \sum_{k,mn} \frac{f_n(k) - f_m(k+q)} {E_n(k) - E_m(k+q)}  \\
	  & & \times \, \left\vert \sum_{\alpha} U_{\alpha \downarrow}^{m\ast} (k+q) U_{\alpha \uparrow}^n (k) \right\vert ^2, \nonumber\\
	B_2(q) & = & \frac{J_\mathrm{H}^2 }{2S} \sum_{k,mn} \frac{f_n(k) - f_m(k+q)} {E_n(k) - E_m(k+q)}  \\
	& &  \times \, \sum_{\alpha \beta} U_{\alpha \downarrow}^{m\ast} (k+q) U_{\alpha \uparrow}^n (k) U_{\beta \downarrow}^{n\ast} (k) U_{\beta \uparrow}^m (k+q), \nonumber
\end{eqnarray}
where $f_n(k) = 1/(1+e^{\beta (E_n(k)-\mu)})$ denotes the Fermi-distribution function with $\mu$ the chemical potential. The Hamiltonian (\ref{eq:swfinal}) is diagonalized by
a Bogoliubov transformation yielding the spin-wave dispersion
\begin{eqnarray}
\omega(q) & = &  \sqrt{A^2(q)-B^2(q)}
\end{eqnarray}
of the system in the presence of both, super- and double-exchange.

\section{Results}
\label{sec:results}

\subsection{Classical phase diagram}

\begin{figure}[t]
  \centering
  \includegraphics[width=8cm]{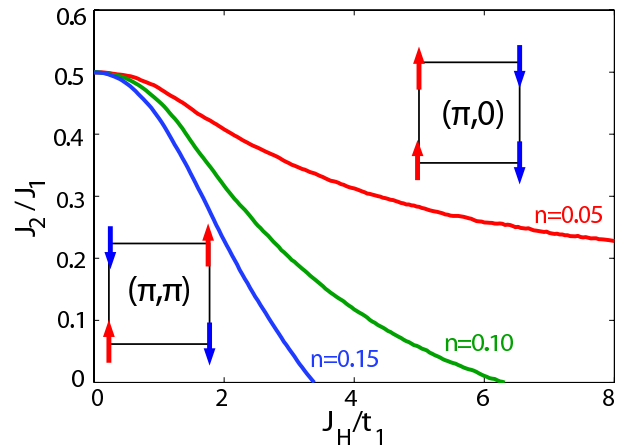}
  \caption{(Color online) Classical phase diagrams of the degenerate double-exchange model as a function of
  $J_\mathrm{H}/t_1$ and $J_2/J_1$ for different filling levels $n=0.05$, $n=0.10$, and $n=0.15$. 
  The tight-binding hopping parameters are $t_2=-0.1t_1$, $t_3=0.2t_1$, and $t_4=0.05t_1$, the nearest-neighbor superexchange $J_1 = 0.04t_1$.}
  \label{fig:phase}
\end{figure}

As a pre-requisite for the spin-wave expansion, we first have to
identify the regime where $(\pi,0)$ stripe antiferromagnetism is
classically stable. When $J_\mathrm{H}=0$ or at zero filling ($n=0$),
$(\pi,0)$ order is stable for $J_2>J_1/2$. In this regime, a finite Hund coupling further stabilizes $(\pi,0)$ order since
the itinerant electrons are more likely to occupy the $d_{yz}$ orbitals which have a larger overlap along the ferromagnetic $y$-direction. Hence the double-exchange 
effectively weakens the spin coupling along the $y$ direction and reduces the magnetic frustration. On the other hand, for $J_2<J_1/2$ the $(\pi,\pi)$ N\'eel antiferromagnet
is the classical ground state for $J_\mathrm{H}\to 0$. However, from the DDEX models for the manganites, it is known\cite{Brink+99a,Brink+99b,Brink+01} that even 
for $J_2=0$, a sufficiently strong Hund coupling will eventually stabilize $(\pi,0)$ order.  

For $J_2<J_1/2$, the phase boundary between the two classical ground states is determined by the condition that the electronic kinetic-energy 
gain for the orbitally-polarized and stripe-ordered configuration equals the difference of the magnetic energies, $E_{(\pi,0)}-E_{(\pi,\pi)}=2J_1-4J_2$. 
The electronic energies for a given electron filling $n$ 
are easily calculated by diagonalizing the free-electron Hamiltonian $\Ham_\mathrm{e}$ (\ref{eq:heq}) for the two different spin configurations. 
From now on, all the energies will be expressed in the unit of $t_1$, the largest of the hopping amplitudes. The tight-binding 
parameters will be chosen to be $t_2=-0.1t_1$, $t_3=0.2t_1$, and $t_4=0.05t_1$ throughout the paper. We further set $J_1=0.04t_1$ so that the exchange constants 
are of the order of 10~meV for a bandwidth of 1~eV, in agreement  with both numerical\cite{Yildirim+08} and experimental\cite{Zhao+09} observations. 

The resulting phase diagrams for electron fillings $n=0.05$, $n=0.1$, and $n=0.15$ are shown in Fig.~\ref{fig:phase}. As predicted, a sufficiently large 
$J_\mathrm{H}$ can stabilize $(\pi,0)$ stripe order even when N\'eel order is favored by $J_2<J_1/2$. Moreover, the $(\pi,0)$ phase is enhanced 
for larger filling $n$ because the kinetic energy gain increases with the number of itinerant electrons. Although similar results have been 
obtained in the context of the manganites,\cite{Brink+99a,Brink+99b,Brink+01} we point out that there are crucial differences. Whereas in the manganites 
$J_2$ is negligible and $(\pi,0)$ antiferromagnetism obtains because of the enormously large
 Hund coupling, in the pnictides, $J_\mathrm{H}$
is much smaller and of the order of the electronic bandwidth.\cite{Haule+08,Craco+08,Nakamura+08,Vildosola+08,Anisimov+08,Yang+09} Therefore, the large $J_2$,
which has been predicted early on\cite{Si+08} based on the geometry of the ion-arsenic layers, is essential for stripe antiferromagnetism in the pnictides.

In the following calculations we will mostly focus on the regime
$J_2>J_1/2$, where $(\pi,0)$ order remains classically stable for
$J_\mathrm{H}\to 0$.  In the calculation, our
primary interest is in how the spin-wave spectrum is renormalized
as we gradually turn on $J_\mathrm{H}$. In principle, when $J_2<J_1/2$, the linear spin-wave calculations can 
still be carried out for a $J_\mathrm{H}$ that is large enough to classically stabilize the stripe order.

\subsection{Orbital and spin polarization}

\begin{figure}[t]
  \centering
  \includegraphics[width=\linewidth]{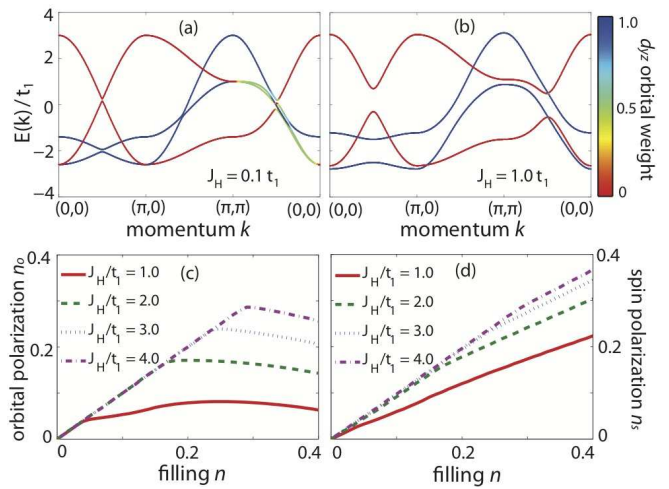}
  \caption{(Color online) (a,~b) The dispersions of the itinerant-
    electron bands for $J_\mathrm{H} = 0.1t_1$(a) and $J_\mathrm{H} =
    1.0t_1$(b), along the path $(0,0)$-$(\pi,0)$-$(\pi,\pi)$-$(0,0)$
    in the Brillouin zone containing one Fe atom per unit cell. The
    tight-binding parameters are chosen to be $t_2=-0.1t_1$, $t_3=0.2t_1$, and $t_4=0.05t_1$ throughout the paper. The bands are colored according to the weight of the $d_{yz}$ orbital. (c,~d) Total orbital polarization $n_o$(c) and spin polarization $n_s$(d) as a function of the filling $n$ for various Hund's couplings $J_\mathrm{H}$.}
  \label{fig:opsp}
\end{figure}

We proceed with a more careful inspection of the electronic Hamiltonian $\Ham_\mathrm{e}$ (\ref{eq:heq}) which includes Hund's coupling to the local moments on a classical 
level given by $\Ham_\mathrm{H}^{(0)}$. Here we focus on the regime where the classical magnetic ground state $\kappa_i = \exp{(i\bm{Q} \cdot \bm{r}_i)}= \pm 1$ is  
the stripe antiferromagnet, $\bm{Q}=(\pi,0)$. Certainly, this is always the case for $J_2>J_1/2$. From the diagonalization of $\Ham_\mathrm{e}$ (\ref{eq:heq}) with the same 
set of hopping parameters used previously, we obtain the dispersions of the electronic bands $E_n(k)$ shown in Fig.~\ref{fig:opsp}(a) and (b) for $J_\mathrm{H} = 0.1t_1$ and 
$J_\mathrm{H} = 1.0t_1$, respectively. Since the diagonal hopping amplitude $t_4$ between different orbitals is non-zero, the bands are always momentum-dependent
superpositions of the two orbitals $d_{yz}$ and $d_{xz}$ as indicated by the color coding. For small $J_\mathrm{H}$ [Fig.~\ref{fig:opsp}(a)], the orbital hybridization is found  
to be strong for some momenta, for example, along the $(0,0)$-$(\pi,\pi)$ direction. However, in the relevant regime where Hund's coupling is of the order of the electronic bandwidth, 
e.g. $J_\mathrm{H} = 1.0t_1$  [Fig.~\ref{fig:opsp}(b)], the gaps between the bands are pronounced and the orbital polarization of each band remains almost perfect for all 
momenta. Therefore, from now on, we will denote these bands as `$d_{xz}$' and `$d_{yz}$' for simplicity.

We further calculate the total orbital polarization $n_o = \sum_\nu(\rho_{yz,\nu}-\rho_{xz,\nu})$ and spin polarization 
$n_s = \sum_\alpha(\rho_{\alpha,\uparrow}-\rho_{\alpha,\downarrow})$. Here, $\rho_{\alpha\nu} = \langle \tilde{c}_{i\alpha\nu}^\dagger \tilde{c}_{i\alpha\nu}\rangle$
is defined as the density of electrons in orbital  $\alpha=xz,yz$ with spin $\nu=\uparrow,\downarrow$. Obviously, these densities sum up to the total filling of the bands, 
$n = \sum_{\alpha\nu} \rho_{\alpha\nu}$. Note that since the densities are defined in the sub-lattice rotated basis,
$\nu=\uparrow$ corresponds to an electron spin aligned with the local moment spin. 

In Fig.~\ref{fig:opsp}(c), the orbital polarization $n_o$ is shown as a function of the total filling $n$ for different values of $J_H$. For small $n$
the electrons populate only the lowest band with almost perfect $d_{yz}$ character and therefore $n_o\simeq n$. Indeed, the slope is very close to one 
indicating that the admixture of the $d_{xz}$ orbital to the lowest
band is negligible and that the orbital polarization can be considered
to be perfect.
The emergent ferro-orbital order, illustrated in Fig. \ref{fig:schematic}, is a consequence of the ferromagnetic Hund's coupling which wants to align the electron 
spin with the local moments and therefore suppresses the electron motion along the antiferromagnetic spin direction. Since $t_1>|t_2|$ the electrons 
populate the $d_{yz}$ orbitals which have a larger overlap along the ferromagnetic $y$ direction. Increasing the filling $n$, the Fermi energy will eventually 
reach the bottom of the next band with mainly $d_{xz}$ character. Above this 
particular filling $\bar{n}$, the orbital polarization is no longer perfect and even starts to decrease with $n$ for larger values of $J_H$.
In general, a larger $J_\mathrm{H}$ increases the energy difference between the two bands, and thus increases the filling $\bar{n}$ up to which the orbital polarization is
perfect. 

Since Hund's coupling tends to align the spins of the itinerant electrons with the local moments, also the spin polarization $n_s$ increases with the
electron filling $n$ and is bigger for larger $J_\mathrm{H}$ as shown
in Fig.~\ref{fig:opsp}(d). However, for small $n$, the slopes are
slightly smaller than
 one, signaling an 
incomplete spin polarization. This is a consequence of the small but finite spin off-diagonal elements in $\Ham_\mathrm{e}$ (\ref{eq:heq}) which do not only depend on 
the smallest hopping amplitude $t_4$, as the orbital hybridization terms, but also on $t_2$ and $t_3$.   For $n>\bar{n}$, the slope of the
spin-polarization curves  is reduced  indicating that the spin polarization of the $d_{yz}$ orbitals is larger than that of the $d_{xz}$ orbitals.

\subsection{Spin-wave spectrum}
\label{sec:sw}

\begin{figure}[t]
  \centering
  \includegraphics[width=8cm]{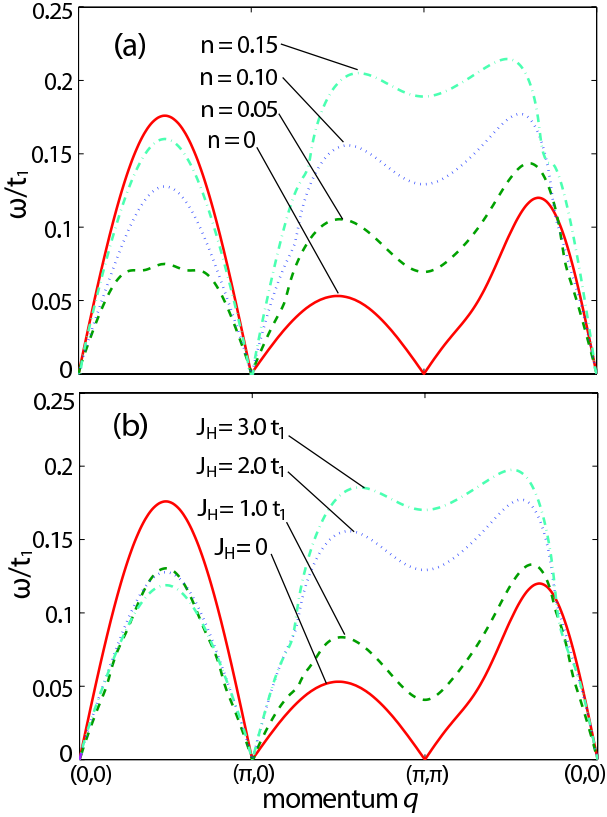}
  \caption{(Color online) Spin-wave dispersion $\omega(q)$ in the presence of both super- and double-exchange along the path $(0,0)$-$(\pi,0)$-$(\pi,\pi)$-$(0,0)$
  for (a) different filling levels $n$, $J_\mathrm{H}=2.0t_1$, and (b) different Hund's couplings $J_\mathrm{H}$, $n=0.1$. In both cases, $J_1=0.04t_1$, $J_2=0.6J_1$, and $S=1/2$.}
  \label{fig:sw}
\end{figure}

Now we set out to calculate the spin-wave spectrum by considering the corrections from the itinerant bands to the $J_1$-$J_2$ Heisenberg model for the 
local moments. The hopping amplitudes are set to the values used before. The spin length $S$ of the local moment is $1/2$, which reflects the relatively small moment measured 
by the experiment\cite{Cruz+08} and is consistent with a local multiplet structure with an orbital degeneracy.\cite{Kruger+09}  Finally, the Heisenberg model is strongly 
frustrated with $J_1 = 0.04t_1$ and $J_2=0.6J_1$.

We first calculate the spin-wave spectra for different filling levels $n$ with $J_\mathrm{H} = 2.0t_1$, shown by Fig.~\ref{fig:sw}(a). In this case, the complete orbital polarization 
is found up to $\bar{n}=0.16$ [see Fig.~\ref{fig:opsp}(c)]. When $n=0$, corresponding to an empty itinerant band, our model reduces to an
isotropic $J_1$-$J_2$ Heisenberg model, where the linear spin-wave energies are zero at both, $(\pi,0)$ and $(\pi,\pi)$. At finite electron densities $n<\bar{n}$,  we observe 
that the spin-wave energy at $(\pi,\pi)$ is pushed to higher values as $n$  increases.  This indicates a stabilization of the stripe antiferromagnetism over the competing N\'eel order. 
We also note a significant mode softening along  the $(0,0)$-$(\pi,0)$ direction at low fillings due to the other finite hopping amplitudes $t_2$, $t_3$, and $t_4$.

For $n >\bar{n}$, not shown in the graph, the spin-wave energy at $(\pi,\pi)$ decreases with $n$. This behavior results from the reduction of the orbital polarization [see Fig.~\ref{fig:opsp}(c)] and hence of the anisotropy, once the itinerant electrons populate the next band with mainly $d_{xz}$ character. 
As the filling level continues to rise, the spin-wave spectrum becomes unstable suggesting that the system may evolve to a different ground state.

Fig.~\ref{fig:sw}(b) shows the spin-wave spectra for different values
of the Hund coupling $J_\mathrm{H}$ and a filling level of $n=0.1$. According to Fig.~\ref{fig:opsp}(c), we have complete orbital polarization for all the $J_\mathrm{H}$'s used in Fig.~\ref{fig:sw}(b), except for $J_\mathrm{H}=1.0t_1$. As expected, a larger $J_\mathrm{H}$ produces stronger corrections to the spin-wave dispersion, especially around $(\pi,\pi)$ where the spin-wave energy almost reaches a maximum. In contrast, the dispersion along $(0,0)$ to $(\pi,0)$ is almost unaffected after $J_\mathrm{H}$ reaches a certain value on the order of the electronic bandwidth.

\subsection{Magnetic anisotropy}

\begin{figure}[t]
  \centering
  \includegraphics[width=8cm]{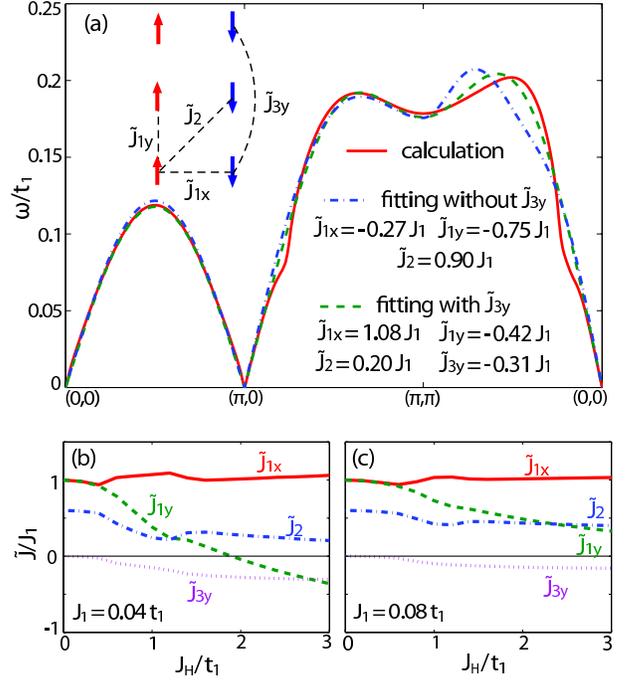}
  \caption{(Color online) (a) Best fits of the spin-wave spectrum of the DDEX for $n=0.1$ and $J_\mathrm{H} = 3.0t_1$ obtained from a spin-only model with effective couplings
  illustrated as inset. (b,~c) Effective exchange constants as functions of Hund's coupling $J_\mathrm{H}$ for bare exchange constants $J_1=0.04t_1$(b), $J_1=0.08t_1$(c), and
  $J_2/J_1=0.6$.}
  \label{fig:fit}
\end{figure}

In Section~\ref{sec:sw} we saw that the double exchange leads to a dramatic change of the magnetic excitation spectra. In particular, the spin-wave modes at the
N\'eel antiferromagnetic wave vector $(\pi,\pi)$ are almost pushed to a maximum consistent with the neutron-scattering data.\cite{Zhao+09} Since orbital ordering 
leads to a dramatic anisotropy in the electronic structure, the ferromagnetic double-exchange contribution is expected to be much larger along the $y$-direction along where the ferromagnetic spin chains are formed. To quantify the induced magnetic anisotropy, in this Section, we fit the spin-wave dispersions calculated in the presence of 
both, super- and double-exchange to an effective, anisotropic Heisenberg model. 

To be more specific, in the effective spin-only model, nearest-neighbors are coupled by exchanges $\tJ_{1x}$ and $\tJ_{1y}$ along the
$x$ and $y$ directions, respectively, and next-nearest neighbors by
$\tJ_2$ [see inset of Fig.~\ref{fig:fit}(a)]. Please note that
different symbols are used here to distinguish 
from the couplings in the original Heisenberg model $\Ham_\mathrm{loc}$ (\ref{eq:heisenberg}). 
Furthermore, we introduce an additional ferromagnetic exchange $\tJ_{3y}<0$ between the third-nearest neighbors along the ferromagnetic chains.  
The utility of retaining longer-ranged ferromagnetic couplings has been demonstrated in the ferromagnetic double-exchange model, 
where simple nearest-neighbor exchange is unable to reproduce the calculated dispersions from either canonical transformations\cite{Nagaev+98,Golosov+00} or diagrammatic perturbation theory.\cite{Shannon+02} Also for manganites with so called CE-type charge-spin-orbital order,  longer-ranged ferromagnetic couplings along the 
ferromagnetic zig-zag chains are crucial in order to obtain a good fit of the magnetic excitation spectra.\cite{Senff+06}
 
In the relevant regime $\tJ_{1x}\ge\tJ_{1y}$, $\tJ_{1y}<2 \tJ_2$, $\tJ_{3y}<0$, where $(\pi,0)$ stripe antiferromagnetism is stable, the linear-spin wave 
dispersion $\tilde{\omega}(q)$ is determined by $\tilde{\omega}^2(q) = \tilde{A}^2(q)-\tilde{B}^2(q)$ with $\tilde{A}(q)=\tJ_{1x}-\tJ_{1y}(1-\cos{q_y})+2\tJ_2 - \tJ_{3y}(1-\cos{2q_y})$ 
and $\tilde{B}(q)=\tJ_{1x}\cos{q_x}+2\tJ_2\cos{q_x}\cos{q_y}$.

In Fig.~\ref{fig:fit}(a), the fitted spin-wave dispersions to the spectrum calculated from the DDEX model with $n=0.1$ and $J_\mathrm{H} = 3.0t_1$
are shown.  Indeed, the inclusion of the longer-ranged ferromagnetic coupling $\tJ_{3y}$ leads to a significant improvement of the fit.
Moreover, setting $\tJ_{3y} = 0$ gives an unrealistically large correction to the nearest-neighbor exchange along the $x$-direction ($\tJ_{1x} = -0.27 J_1$), which has to 
be compensated by a fairly substantial increase in the diagonal exchange, $\tJ_2 = 0.90 J_1$. This result is certainly unphysical since from the classical double-exchange 
argument, we expect $\tJ_{1x}\simeq J_1$ and $\tJ_2<J_2$. In contrast, by including $\tJ_{3y}$, we obtain more physical fitting results. 
Consequently, in the following, all of the fittings will be performed with $\tJ_{3y} \neq 0$.

\begin{figure}[t]
  \centering
  \includegraphics[width=8cm]{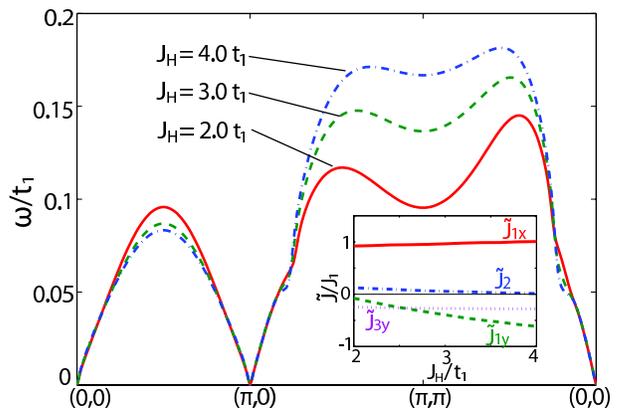}
  \caption{(Color online) Spin-wave dispersion $\omega(q)$ in the presence of both, super- and double-exchange along the path $(0,0)$-$(\pi,0)$-$(\pi,\pi)$-$(0,0)$
  for different Hund's couplings $J_\mathrm{H}$.  Parameters are the same as in Fig.~\ref{fig:sw}(b), except for $J_2/J_1=0.4$. In this regime, $(\pi,0)$ order is classically 
  stable for $J_\mathrm{H}\gtrsim 1.6t_1$. (Inset: the fitted, effective exchange constants $\tJ_{1x}$, $\tJ_{1y}$, $\tJ_2$, and $\tJ_{3y}$ as functions of $J_\mathrm{H}$.)}
  \label{fig:more}
\end{figure}

The resulting effective exchange constants as a function of Hund's coupling $J_\mathrm{H}$ are shown in Fig.~\ref{fig:fit}(b) and (c) 
for $J_1 = 0.04t_1$ and $J_1 = 0.08t_1$, respectively. Note that for $J_1 = 0.04t_1$, the spectra of the DDEX model are shown in Fig.~\ref{fig:sw}(b).
In both cases we use $n=0.1$, $J_2/J_1=0.6$, and the same tight-binding parameters as before. 
Obviously, for $J_\mathrm{H}=0$ the effective exchange constants have to be identical with the bare ones, $\tJ_{1x}=\tJ_{1y}=J_1$, $\tJ_2=J_2$, and $\tJ_3=0$.
Increasing Hund's coupling does not change $\tJ_{1x}$, whereas the other exchanges decrease due to the different ferromagnetic double-exchange 
contributions. Since $t_1$ is the largest hopping amplitude and most electrons populate the $d_{yx}$ orbitals, the fastest decrease is observed for the 
coupling $\tJ_{1y}$ along the ferromagnetic chains. We also note that the non-monotonic behavior for small $J_\mathrm{H}$ is not unphysical, but due to the fact that 
for $n=0.1$, complete orbital polarization is not achieved until $J_\mathrm{H} \gtrsim 1.5t_1$. 

Interestingly, in the case $J_1 = 0.04t_1$ [see Fig.~\ref{fig:fit}(b)],  the coupling $\tJ_{1y}$ along the ferromagnetic direction becomes negative for $J_\mathrm{H} \gtrsim 2.0t_1$,
which completely removes the frustration in the effective spin-only model. Such an effective negative exchange coupling along the ferromagnetic spin direction has been used phenomenologically to rationalize the spectra measured by inelastic neutron scattering.\cite{Zhao+09}
Remarkably, around $J_\mathrm{H} \approx 2.0t_1$, the ratios of the three exchange constants $\tJ_{1x}$, $\tJ_{1y}$, and $\tJ_{2}$, agree extremely well with 
the experimental estimates.\cite{Zhao+09} As shown in Fig.~\ref{fig:fit}(c), for $J_1 = 0.08t_1$, the relative corrections to the exchange couplings are too small to make 
$\tJ_{1y}$ ferromagnetic for realistic values of $J_\mathrm{H}$. In order to achieve the experimentally observed negative $\tJ_{1y}$ within reasonable parameter space, 
we require $J_1\lesssim 0.05t_1$. In fact, inelastic neutron scattering\cite{Zhao+09} suggests that the exchange constants are of the order of 10~meV, which in our 
theory leads to an electronic bandwidth and  Hund's coupling $J_\mathrm{H}$ both of the order of 1~eV, in agreement with other experimental estimates.\cite{Yang+09}

Though the parameter regime $J_2/J_1<1/2$ is most likely not relevant
to the pnictides, we consider it nonetheless for completeness.  In
this regime, the Hund coupling has to exceed a critical value in order to stabilize 
$(\pi,0)$ stripe antiferromagnetism.  As an example, we assume $J_2 = 0.4 J_1$ and $n=0.1$, in which case $(\pi,0)$ order is classically stable for $J_\mathrm{H}\gtrsim 1.6t_1$
[see Fig.~\ref{fig:phase}]. The resulting spin-wave spectra are shown
in Fig.~\ref{fig:more} for different values of $J_\mathrm{H}$. The effective exchange couplings obtained 
from fitting to the spin-only model are shown in the inset. Although the behavior is qualitatively similar to the case $J_2/J_1=0.6$, the agreement with the experimental
data is not as good. Moreover, in order to push the dispersion at $(\pi,\pi)$ close to a maximum, we need a much bigger $J_\mathrm{H}$ which can not be justified for 
the pnictides.

\section{Discussion and conclusion}
\label{sec:disc}

In summary, we have studied a DDEX model for the iron pnictides which explains the dramatic magnetic\cite{Zhao+08,Zhao+09}
and electronic\cite{Chuang+10,Chu+10} anisotropies in these materials. The model consists of local moments which are coupled by antiferromagnetic superexchanges
$J_1$ and $J_2$ between nearest- and next-nearest-neighbor spins, respectively, and of itinerant electrons in the bands of the degenerate $d_{xz}$ and $d_{yz}$ orbitals.
The electrons are coupled  to the local moments by a ferromagnetic Hund exchange, $J_H$. The system spontaneously develops ferro-orbital order because of the
kinetic energy gained by directing the itinerant electrons along ferromagnetic spin chains which are stabilized by the double-exchange mechanism. 

Although similar results have been obtained previously in the context of the manganites,\cite{Brink+99a,Brink+99b,Brink+01} we point out that there is a crucial difference 
between the two classes of materials. Whereas in the pnictides, the local moments are inherently quantum ($S=1/2$) and $J_\mathrm{H}$ is at most of the order of the 
electronic bandwidth, the DDEX models for the manganites can be greatly simplified by assuming an infinitely large $J_\mathrm{H}$ and by treating the core spins as classical.
Because of these approximations, the spin-wave excitations of the ferro-orbitally ordered stripe antiferromagnet have not been addressed in the manganite literature. 

In this work, we have explicitly taken the quantum nature of the local moments into account and calculated the magnetic excitation spectra in the presence
 of both, super- and double-exchange. Over a realistic parameter range, the calculated spin-wave dispersions are found to be in good agreement with the neutron-scattering
data.\cite{Zhao+08,Zhao+09} In particular, we find that the dispersion is pushed almost to a local maximum at the competing N\'eel ordering wave vector as
seen in the experiment.\cite{Zhao+09} By fitting to an effective spin-only model, we find that the coupling along the ferromagnetic direction becomes negative, 
which demonstrates that in the pnictides, the double-exchange along the ferromagnetic spin-direction can overcompensate the 
superexchange between the local moments. In this regime, $(\pi,0)$ antiferromagnetism is unfrustrated.

It is feasible that the parent, undoped materials self-tune the size of the local moments and the carrier density to the point where the $(\pi,0)$ antiferromagnetism is
most stable. In our theory, this is the case for the optimal filling level $n=\bar{n}$ where the orbital polarization reaches a maximum.  In fact, the starting Heisenberg 
model $\Ham_\mathrm{loc}$ (\ref{eq:heisenberg}) is likely to be in the regime of strong frustration, $0.4<J_2/J_1<0.6$, where quantum fluctuations destroy long
range magnetic order. Only through the interaction with the itinerant
electrons does stripe antiferromagnetism emerge. Electron- or hole-doping the system 
at $n=\bar{n}$ diminishes the orbital order and thus increases the magnetic frustration,  lowering the transition temperature to a spin-ordered state.

We conclude by stressing that the degenerate double-exchange model motivated and studied in this work qualitatively explains many properties of the undoped and
slightly doped iron pnictides. The emergent ferro-orbital order breaks the in-plane lattice symmetry, thereby driving the orthorhombic lattice distortion. Further,
orbital order induces a strong electronic anisotropy which explains why the structural transition is accompanied by dramatic Fermi-surface
reconstructions\cite{Shimojima+10} and transport anomalies.\cite{McGuire+08} Moreover, recent STM experiments\cite{Chuang+10} and in-plane resistivity 
measurements\cite{Tanatar+10,Chu+10} have unambiguously demonstrated the spatial anisotropy of the electronic state at temperatures below the
structural transition temperature. Finally, orbital ordering  produces strong magnetic anisotropy,
essential for explaining the experimentally observed magnetic excitation spectra.\cite{Zhao+08,Zhao+09}

During the review of this work, several mean-field calculations\cite{Bascones+10,Daghofer+10,Chen+10} based on five-orbital Hubbard models have shown that 
the total occupations of the $d_{xz}$ and $d_{yz}$ orbitals are close to each other, but that there is a significant difference in the density of states for the two orbitals 
at the chemical potential. The ferro-orbital order in our theory does not contradict these results. In fact, by viewing the itinerant electrons as residual quasiparticles on 
top of local moments formed because of strong correlations, our model is an effective low-energy theory that successfully describes the strongly anisotropic 
electron state around the Fermi surface of the iron pnictides.

\begin{acknowledgments}
This work is partially funded by NSF under Grant No.~DMR-0940992. The authors acknowledge support from the Center for
Emergent Superconductivity, an Energy Frontier Research Center funded by
the U.S. Department of Energy, Office of Science, Office of Basic Energy
Sciences under Award Number DE-AC0298CH1088.
\end{acknowledgments}


\begin{thebibliography}{74}
\expandafter\ifx\csname natexlab\endcsname\relax\def\natexlab#1{#1}\fi
\expandafter\ifx\csname bibnamefont\endcsname\relax
  \def\bibnamefont#1{#1}\fi
\expandafter\ifx\csname bibfnamefont\endcsname\relax
  \def\bibfnamefont#1{#1}\fi
\expandafter\ifx\csname citenamefont\endcsname\relax
  \def\citenamefont#1{#1}\fi
\expandafter\ifx\csname url\endcsname\relax
  \def\url#1{\texttt{#1}}\fi
\expandafter\ifx\csname urlprefix\endcsname\relax\def\urlprefix{URL }\fi
\providecommand{\bibinfo}[2]{#2}
\providecommand{\eprint}[2][]{\url{#2}}

\bibitem[{\citenamefont{Kamihara et~al.}(2008)\citenamefont{Kamihara, Watanabe,
  Hirano, and Hosono}}]{Kamihara+08}
\bibinfo{author}{\bibfnamefont{Y.}~\bibnamefont{Kamihara}},
  \bibinfo{author}{\bibfnamefont{T.}~\bibnamefont{Watanabe}},
  \bibinfo{author}{\bibfnamefont{M.}~\bibnamefont{Hirano}}, \bibnamefont{and}
  \bibinfo{author}{\bibfnamefont{H.}~\bibnamefont{Hosono}},
  \bibinfo{journal}{J. Am. Chem. Soc.} \textbf{\bibinfo{volume}{130}},
  \bibinfo{pages}{3296} (\bibinfo{year}{2008}).

\bibitem[{\citenamefont{Chen et~al.}(2008{\natexlab{a}})\citenamefont{Chen, Wu,
  Wu, Liu, Chen, and Fang}}]{Chen+08a}
\bibinfo{author}{\bibfnamefont{X.~H.} \bibnamefont{Chen}},
  \bibinfo{author}{\bibfnamefont{T.}~\bibnamefont{Wu}},
  \bibinfo{author}{\bibfnamefont{G.}~\bibnamefont{Wu}},
  \bibinfo{author}{\bibfnamefont{R.~H.} \bibnamefont{Liu}},
  \bibinfo{author}{\bibfnamefont{H.}~\bibnamefont{Chen}}, \bibnamefont{and}
  \bibinfo{author}{\bibfnamefont{D.~F.} \bibnamefont{Fang}},
  \bibinfo{journal}{Nature (London)} \textbf{\bibinfo{volume}{453}},
  \bibinfo{pages}{761} (\bibinfo{year}{2008}{\natexlab{a}}).

\bibitem[{\citenamefont{Chen et~al.}(2008{\natexlab{b}})\citenamefont{Chen, Li,
  Wu, Li, Hu, Dong, Zheng, Luo, and Wang}}]{Chen+08b}
\bibinfo{author}{\bibfnamefont{G.~F.} \bibnamefont{Chen}},
  \bibinfo{author}{\bibfnamefont{Z.}~\bibnamefont{Li}},
  \bibinfo{author}{\bibfnamefont{D.}~\bibnamefont{Wu}},
  \bibinfo{author}{\bibfnamefont{G.}~\bibnamefont{Li}},
  \bibinfo{author}{\bibfnamefont{W.~Z.} \bibnamefont{Hu}},
  \bibinfo{author}{\bibfnamefont{J.}~\bibnamefont{Dong}},
  \bibinfo{author}{\bibfnamefont{P.}~\bibnamefont{Zheng}},
  \bibinfo{author}{\bibfnamefont{J.~L.} \bibnamefont{Luo}}, \bibnamefont{and}
  \bibinfo{author}{\bibfnamefont{N.~L.} \bibnamefont{Wang}},
  \bibinfo{journal}{Phys. Rev. Lett.} \textbf{\bibinfo{volume}{100}},
  \bibinfo{pages}{247002} (\bibinfo{year}{2008}{\natexlab{b}}).

\bibitem[{\citenamefont{Ren et~al.}(2008)\citenamefont{Ren, Che, Dong, Yang,
  Lu, Yi, Shen, Li, Sun, Zhou et~al.}}]{Ren+08}
\bibinfo{author}{\bibfnamefont{Z.~A.} \bibnamefont{Ren}},
  \bibinfo{author}{\bibfnamefont{G.~C.} \bibnamefont{Che}},
  \bibinfo{author}{\bibfnamefont{X.~L.} \bibnamefont{Dong}},
  \bibinfo{author}{\bibfnamefont{J.}~\bibnamefont{Yang}},
  \bibinfo{author}{\bibfnamefont{W.}~\bibnamefont{Lu}},
  \bibinfo{author}{\bibfnamefont{W.}~\bibnamefont{Yi}},
  \bibinfo{author}{\bibfnamefont{X.~L.} \bibnamefont{Shen}},
  \bibinfo{author}{\bibfnamefont{Z.~C.} \bibnamefont{Li}},
  \bibinfo{author}{\bibfnamefont{L.~L.} \bibnamefont{Sun}},
  \bibinfo{author}{\bibfnamefont{F.}~\bibnamefont{Zhou}}, \bibnamefont{et~al.},
  \bibinfo{journal}{Europhys. Lett.} \textbf{\bibinfo{volume}{83}},
  \bibinfo{pages}{17002} (\bibinfo{year}{2008}).

\bibitem[{\citenamefont{Wen et~al.}(2008)\citenamefont{Wen, Mu, Fang, Yang, and
  Zhu}}]{Wen+08}
\bibinfo{author}{\bibfnamefont{H.~H.} \bibnamefont{Wen}},
  \bibinfo{author}{\bibfnamefont{G.}~\bibnamefont{Mu}},
  \bibinfo{author}{\bibfnamefont{L.}~\bibnamefont{Fang}},
  \bibinfo{author}{\bibfnamefont{H.}~\bibnamefont{Yang}}, \bibnamefont{and}
  \bibinfo{author}{\bibfnamefont{X.~Y.} \bibnamefont{Zhu}},
  \bibinfo{journal}{Europhys. Lett.} \textbf{\bibinfo{volume}{82}},
  \bibinfo{pages}{17009} (\bibinfo{year}{2008}).

\bibitem[{\citenamefont{Rotter et~al.}(2008)\citenamefont{Rotter, Tegel, and
  Johrendt}}]{Rotter+08}
\bibinfo{author}{\bibfnamefont{M.}~\bibnamefont{Rotter}},
  \bibinfo{author}{\bibfnamefont{M.}~\bibnamefont{Tegel}}, \bibnamefont{and}
  \bibinfo{author}{\bibfnamefont{D.}~\bibnamefont{Johrendt}},
  \bibinfo{journal}{Phys. Rev. Lett.} \textbf{\bibinfo{volume}{101}},
  \bibinfo{pages}{107006} (\bibinfo{year}{2008}).

\bibitem[{\citenamefont{Kivelson and Yao}(2008)}]{Kivelson+08}
\bibinfo{author}{\bibfnamefont{S.~A.} \bibnamefont{Kivelson}} \bibnamefont{and}
  \bibinfo{author}{\bibfnamefont{H.}~\bibnamefont{Yao}},
  \bibinfo{journal}{Nature Materials} \textbf{\bibinfo{volume}{7}},
  \bibinfo{pages}{927} (\bibinfo{year}{2008}).

\bibitem[{\citenamefont{Zaanen}(2009)}]{Zaanen09}
\bibinfo{author}{\bibfnamefont{J.}~\bibnamefont{Zaanen}},
  \bibinfo{journal}{Nature} \textbf{\bibinfo{volume}{457}},
  \bibinfo{pages}{546} (\bibinfo{year}{2009}).

\bibitem[{\citenamefont{Tesanovic}(2009)}]{Tesanovic09}
\bibinfo{author}{\bibfnamefont{S.}~\bibnamefont{Tesanovic}},
  \bibinfo{journal}{Physics} \textbf{\bibinfo{volume}{2}}, \bibinfo{pages}{60}
  (\bibinfo{year}{2009}).

\bibitem[{\citenamefont{de~la Cruz et~al.}(2008)\citenamefont{de~la Cruz,
  Huang, Lynn, J.~Li, Zarestky, Mook, Chen, Luo, Wang, and Dai}}]{Cruz+08}
\bibinfo{author}{\bibfnamefont{C.}~\bibnamefont{de~la Cruz}},
  \bibinfo{author}{\bibfnamefont{Q.}~\bibnamefont{Huang}},
  \bibinfo{author}{\bibfnamefont{J.~W.} \bibnamefont{Lynn}},
  \bibinfo{author}{\bibfnamefont{W.~R.} \bibnamefont{J.~Li}},
  \bibinfo{author}{\bibfnamefont{J.~L.} \bibnamefont{Zarestky}},
  \bibinfo{author}{\bibfnamefont{H.~A.} \bibnamefont{Mook}},
  \bibinfo{author}{\bibfnamefont{G.~F.} \bibnamefont{Chen}},
  \bibinfo{author}{\bibfnamefont{J.~L.} \bibnamefont{Luo}},
  \bibinfo{author}{\bibfnamefont{N.~L.} \bibnamefont{Wang}}, \bibnamefont{and}
  \bibinfo{author}{\bibfnamefont{P.}~\bibnamefont{Dai}},
  \bibinfo{journal}{Nature (London)} \textbf{\bibinfo{volume}{453}},
  \bibinfo{pages}{899} (\bibinfo{year}{2008}).

\bibitem[{\citenamefont{Zhao et~al.}(2008{\natexlab{a}})\citenamefont{Zhao,
  Huang, de~la Cruz, Li, Lynn, Chen, Green, Chen, Li, Li et~al.}}]{Zhao+08b}
\bibinfo{author}{\bibfnamefont{J.}~\bibnamefont{Zhao}},
  \bibinfo{author}{\bibfnamefont{Q.}~\bibnamefont{Huang}},
  \bibinfo{author}{\bibfnamefont{C.}~\bibnamefont{de~la Cruz}},
  \bibinfo{author}{\bibfnamefont{S.}~\bibnamefont{Li}},
  \bibinfo{author}{\bibfnamefont{J.~W.} \bibnamefont{Lynn}},
  \bibinfo{author}{\bibfnamefont{Y.}~\bibnamefont{Chen}},
  \bibinfo{author}{\bibfnamefont{M.~A.} \bibnamefont{Green}},
  \bibinfo{author}{\bibfnamefont{G.~F.} \bibnamefont{Chen}},
  \bibinfo{author}{\bibfnamefont{G.}~\bibnamefont{Li}},
  \bibinfo{author}{\bibfnamefont{Y.}~\bibnamefont{Li}}, \bibnamefont{et~al.},
  \bibinfo{journal}{Nature Materials} \textbf{\bibinfo{volume}{7}},
  \bibinfo{pages}{953} (\bibinfo{year}{2008}{\natexlab{a}}).

\bibitem[{\citenamefont{Zhao et~al.}(2008{\natexlab{b}})\citenamefont{Zhao,
  Huang, de~la Cruz, Lynn, Lumsden, Ren, Yang, Shen, Dong, Zhao
  et~al.}}]{Zhao+08c}
\bibinfo{author}{\bibfnamefont{J.}~\bibnamefont{Zhao}},
  \bibinfo{author}{\bibfnamefont{Q.}~\bibnamefont{Huang}},
  \bibinfo{author}{\bibfnamefont{C.}~\bibnamefont{de~la Cruz}},
  \bibinfo{author}{\bibfnamefont{J.~W.} \bibnamefont{Lynn}},
  \bibinfo{author}{\bibfnamefont{M.~D.} \bibnamefont{Lumsden}},
  \bibinfo{author}{\bibfnamefont{Z.~A.} \bibnamefont{Ren}},
  \bibinfo{author}{\bibfnamefont{J.}~\bibnamefont{Yang}},
  \bibinfo{author}{\bibfnamefont{X.}~\bibnamefont{Shen}},
  \bibinfo{author}{\bibfnamefont{X.}~\bibnamefont{Dong}},
  \bibinfo{author}{\bibfnamefont{Z.}~\bibnamefont{Zhao}}, \bibnamefont{et~al.},
  \bibinfo{journal}{Phys. Rev. B} \textbf{\bibinfo{volume}{78}},
  \bibinfo{pages}{132504} (\bibinfo{year}{2008}{\natexlab{b}}).

\bibitem[{\citenamefont{Kimber et~al.}(2008)\citenamefont{Kimber, Argyriou,
  Yokaichiya, Habicht, Gerischer, Hansen, Chatterji, Klingeler, Hess, Behr
  et~al.}}]{Kimber+08}
\bibinfo{author}{\bibfnamefont{S.~A.~J.} \bibnamefont{Kimber}},
  \bibinfo{author}{\bibfnamefont{D.~N.} \bibnamefont{Argyriou}},
  \bibinfo{author}{\bibfnamefont{F.}~\bibnamefont{Yokaichiya}},
  \bibinfo{author}{\bibfnamefont{K.}~\bibnamefont{Habicht}},
  \bibinfo{author}{\bibfnamefont{S.}~\bibnamefont{Gerischer}},
  \bibinfo{author}{\bibfnamefont{T.}~\bibnamefont{Hansen}},
  \bibinfo{author}{\bibfnamefont{T.}~\bibnamefont{Chatterji}},
  \bibinfo{author}{\bibfnamefont{R.}~\bibnamefont{Klingeler}},
  \bibinfo{author}{\bibfnamefont{C.}~\bibnamefont{Hess}},
  \bibinfo{author}{\bibfnamefont{G.}~\bibnamefont{Behr}}, \bibnamefont{et~al.},
  \bibinfo{journal}{Phys. Rev. B} \textbf{\bibinfo{volume}{78}},
  \bibinfo{pages}{140503(R)} (\bibinfo{year}{2008}).

\bibitem[{\citenamefont{Goldman et~al.}(2008)\citenamefont{Goldman, Argyriou,
  Ouladdiaf, Chatterji, Kreyssig, Nandi, Ni, Bud'ko, Canfield, and
  McQueeney}}]{Goldman+08}
\bibinfo{author}{\bibfnamefont{A.~I.} \bibnamefont{Goldman}},
  \bibinfo{author}{\bibfnamefont{D.~N.} \bibnamefont{Argyriou}},
  \bibinfo{author}{\bibfnamefont{B.}~\bibnamefont{Ouladdiaf}},
  \bibinfo{author}{\bibfnamefont{T.}~\bibnamefont{Chatterji}},
  \bibinfo{author}{\bibfnamefont{A.}~\bibnamefont{Kreyssig}},
  \bibinfo{author}{\bibfnamefont{S.}~\bibnamefont{Nandi}},
  \bibinfo{author}{\bibfnamefont{N.}~\bibnamefont{Ni}},
  \bibinfo{author}{\bibfnamefont{S.~L.} \bibnamefont{Bud'ko}},
  \bibinfo{author}{\bibfnamefont{P.~C.} \bibnamefont{Canfield}},
  \bibnamefont{and} \bibinfo{author}{\bibfnamefont{R.~J.}
  \bibnamefont{McQueeney}}, \bibinfo{journal}{Phys. Rev. B}
  \textbf{\bibinfo{volume}{78}}, \bibinfo{pages}{100506(R)}
  (\bibinfo{year}{2008}).

\bibitem[{\citenamefont{Zhao et~al.}(2008{\natexlab{c}})\citenamefont{Zhao,
  Ratcliff, Lynn, Chen, Luo, Wang, Hu, and Dai}}]{Zhao+08d}
\bibinfo{author}{\bibfnamefont{J.}~\bibnamefont{Zhao}},
  \bibinfo{author}{\bibfnamefont{W.}~\bibnamefont{Ratcliff}},
  \bibinfo{author}{\bibfnamefont{J.~W.} \bibnamefont{Lynn}},
  \bibinfo{author}{\bibfnamefont{G.~F.} \bibnamefont{Chen}},
  \bibinfo{author}{\bibfnamefont{J.~L.} \bibnamefont{Luo}},
  \bibinfo{author}{\bibfnamefont{N.~L.} \bibnamefont{Wang}},
  \bibinfo{author}{\bibfnamefont{J.}~\bibnamefont{Hu}}, \bibnamefont{and}
  \bibinfo{author}{\bibfnamefont{P.}~\bibnamefont{Dai}},
  \bibinfo{journal}{Phys. Rev. B} \textbf{\bibinfo{volume}{78}},
  \bibinfo{pages}{140504(R)} (\bibinfo{year}{2008}{\natexlab{c}}).

\bibitem[{\citenamefont{Huang et~al.}(2008)\citenamefont{Huang, Qiu, Bao,
  Green, Lynn, Gasparovic, Wu, Wu, and Chen}}]{Huang+08}
\bibinfo{author}{\bibfnamefont{Q.}~\bibnamefont{Huang}},
  \bibinfo{author}{\bibfnamefont{Y.}~\bibnamefont{Qiu}},
  \bibinfo{author}{\bibfnamefont{W.}~\bibnamefont{Bao}},
  \bibinfo{author}{\bibfnamefont{M.~A.} \bibnamefont{Green}},
  \bibinfo{author}{\bibfnamefont{J.~W.} \bibnamefont{Lynn}},
  \bibinfo{author}{\bibfnamefont{Y.~C.} \bibnamefont{Gasparovic}},
  \bibinfo{author}{\bibfnamefont{T.}~\bibnamefont{Wu}},
  \bibinfo{author}{\bibfnamefont{G.}~\bibnamefont{Wu}}, \bibnamefont{and}
  \bibinfo{author}{\bibfnamefont{X.~H.} \bibnamefont{Chen}},
  \bibinfo{journal}{Phys. Rev. Lett.} \textbf{\bibinfo{volume}{101}},
  \bibinfo{pages}{257003} (\bibinfo{year}{2008}).

\bibitem[{\citenamefont{Wilson et~al.}(2009)\citenamefont{Wilson, Yamani,
  Rotundu, Freelon, Bourret-Courchesne, and Birgeneau}}]{Wilson+09}
\bibinfo{author}{\bibfnamefont{S.~D.} \bibnamefont{Wilson}},
  \bibinfo{author}{\bibfnamefont{Z.}~\bibnamefont{Yamani}},
  \bibinfo{author}{\bibfnamefont{C.~R.} \bibnamefont{Rotundu}},
  \bibinfo{author}{\bibfnamefont{B.}~\bibnamefont{Freelon}},
  \bibinfo{author}{\bibfnamefont{E.}~\bibnamefont{Bourret-Courchesne}},
  \bibnamefont{and} \bibinfo{author}{\bibfnamefont{R.~J.}
  \bibnamefont{Birgeneau}}, \bibinfo{journal}{Phys. Rev. B}
  \textbf{\bibinfo{volume}{79}}, \bibinfo{pages}{184519}
  (\bibinfo{year}{2009}).

\bibitem[{\citenamefont{Chen et~al.}(2008{\natexlab{c}})\citenamefont{Chen,
  Lynn, Li, Li, Chen, Luo, Wang, Dai, dela Cruz, and Mook}}]{Chen+08c}
\bibinfo{author}{\bibfnamefont{Y.}~\bibnamefont{Chen}},
  \bibinfo{author}{\bibfnamefont{J.~W.} \bibnamefont{Lynn}},
  \bibinfo{author}{\bibfnamefont{J.}~\bibnamefont{Li}},
  \bibinfo{author}{\bibfnamefont{G.}~\bibnamefont{Li}},
  \bibinfo{author}{\bibfnamefont{G.~F.} \bibnamefont{Chen}},
  \bibinfo{author}{\bibfnamefont{J.~L.} \bibnamefont{Luo}},
  \bibinfo{author}{\bibfnamefont{N.~L.} \bibnamefont{Wang}},
  \bibinfo{author}{\bibfnamefont{P.}~\bibnamefont{Dai}},
  \bibinfo{author}{\bibfnamefont{C.}~\bibnamefont{dela Cruz}},
  \bibnamefont{and} \bibinfo{author}{\bibfnamefont{H.~A.} \bibnamefont{Mook}},
  \bibinfo{journal}{Phys. Rev. B} \textbf{\bibinfo{volume}{78}},
  \bibinfo{pages}{064515} (\bibinfo{year}{2008}{\natexlab{c}}).

\bibitem[{\citenamefont{Dong et~al.}(2008)\citenamefont{Dong, Zhang, Xu, Li,
  Li, Hu, Wu, Chen, Dai, Luo et~al.}}]{Dong+08}
\bibinfo{author}{\bibfnamefont{J.}~\bibnamefont{Dong}},
  \bibinfo{author}{\bibfnamefont{H.~J.} \bibnamefont{Zhang}},
  \bibinfo{author}{\bibfnamefont{G.}~\bibnamefont{Xu}},
  \bibinfo{author}{\bibfnamefont{Z.}~\bibnamefont{Li}},
  \bibinfo{author}{\bibfnamefont{G.}~\bibnamefont{Li}},
  \bibinfo{author}{\bibfnamefont{W.~Z.} \bibnamefont{Hu}},
  \bibinfo{author}{\bibfnamefont{D.}~\bibnamefont{Wu}},
  \bibinfo{author}{\bibfnamefont{G.~F.} \bibnamefont{Chen}},
  \bibinfo{author}{\bibfnamefont{X.}~\bibnamefont{Dai}},
  \bibinfo{author}{\bibfnamefont{J.~L.} \bibnamefont{Luo}},
  \bibnamefont{et~al.}, \bibinfo{journal}{Europhys. Lett.}
  \textbf{\bibinfo{volume}{83}}, \bibinfo{pages}{27006} (\bibinfo{year}{2008}).

\bibitem[{\citenamefont{Wu et~al.}(2008)\citenamefont{Wu, Phillips, and
  Neto}}]{Wu+08}
\bibinfo{author}{\bibfnamefont{J.}~\bibnamefont{Wu}},
  \bibinfo{author}{\bibfnamefont{P.}~\bibnamefont{Phillips}}, \bibnamefont{and}
  \bibinfo{author}{\bibfnamefont{A.~H.~C.} \bibnamefont{Neto}},
  \bibinfo{journal}{Phys. Rev. Lett.} \textbf{\bibinfo{volume}{101}},
  \bibinfo{pages}{126401} (\bibinfo{year}{2008}).

\bibitem[{\citenamefont{Haule et~al.}(2008)\citenamefont{Haule, Shim, and
  Kotliar}}]{Haule+08}
\bibinfo{author}{\bibfnamefont{K.}~\bibnamefont{Haule}},
  \bibinfo{author}{\bibfnamefont{J.~H.} \bibnamefont{Shim}}, \bibnamefont{and}
  \bibinfo{author}{\bibfnamefont{G.}~\bibnamefont{Kotliar}},
  \bibinfo{journal}{Phys. Rev. Lett.} \textbf{\bibinfo{volume}{100}},
  \bibinfo{pages}{226402} (\bibinfo{year}{2008}).

\bibitem[{\citenamefont{Craco et~al.}(2008)\citenamefont{Craco, Laad, Leoni,
  and Rosner}}]{Craco+08}
\bibinfo{author}{\bibfnamefont{L.}~\bibnamefont{Craco}},
  \bibinfo{author}{\bibfnamefont{M.~S.} \bibnamefont{Laad}},
  \bibinfo{author}{\bibfnamefont{S.}~\bibnamefont{Leoni}}, \bibnamefont{and}
  \bibinfo{author}{\bibfnamefont{H.}~\bibnamefont{Rosner}},
  \bibinfo{journal}{Physical Review B} \textbf{\bibinfo{volume}{78}},
  \bibinfo{pages}{134511} (\bibinfo{year}{2008}).

\bibitem[{\citenamefont{Nakamura et~al.}(2008)\citenamefont{Nakamura, Arita,
  and Imada}}]{Nakamura+08}
\bibinfo{author}{\bibfnamefont{K.}~\bibnamefont{Nakamura}},
  \bibinfo{author}{\bibfnamefont{R.}~\bibnamefont{Arita}}, \bibnamefont{and}
  \bibinfo{author}{\bibfnamefont{M.}~\bibnamefont{Imada}}, \bibinfo{journal}{J.
  Phys. Soc. Japan} \textbf{\bibinfo{volume}{77}}, \bibinfo{pages}{093711}
  (\bibinfo{year}{2008}).

\bibitem[{\citenamefont{Vildosolda et~al.}(2008)\citenamefont{Vildosolda,
  Pourovskii, Arita, Biermann, and Georges}}]{Vildosola+08}
\bibinfo{author}{\bibfnamefont{V.}~\bibnamefont{Vildosolda}},
  \bibinfo{author}{\bibfnamefont{L.}~\bibnamefont{Pourovskii}},
  \bibinfo{author}{\bibfnamefont{R.}~\bibnamefont{Arita}},
  \bibinfo{author}{\bibfnamefont{S.}~\bibnamefont{Biermann}}, \bibnamefont{and}
  \bibinfo{author}{\bibfnamefont{A.}~\bibnamefont{Georges}},
  \bibinfo{journal}{Phys. Rev. B} \textbf{\bibinfo{volume}{78}},
  \bibinfo{pages}{064518} (\bibinfo{year}{2008}).

\bibitem[{\citenamefont{Anisimov et~al.}(2008)\citenamefont{Anisimov, Korotin,
  Streltsov, Kozhevnikov, Kunes, Shorikov, and Korotin}}]{Anisimov+08}
\bibinfo{author}{\bibfnamefont{V.~I.} \bibnamefont{Anisimov}},
  \bibinfo{author}{\bibfnamefont{D.~M.} \bibnamefont{Korotin}},
  \bibinfo{author}{\bibfnamefont{S.~V.} \bibnamefont{Streltsov}},
  \bibinfo{author}{\bibfnamefont{A.~V.} \bibnamefont{Kozhevnikov}},
  \bibinfo{author}{\bibfnamefont{J.}~\bibnamefont{Kunes}},
  \bibinfo{author}{\bibfnamefont{A.~O.} \bibnamefont{Shorikov}},
  \bibnamefont{and} \bibinfo{author}{\bibfnamefont{M.~A.}
  \bibnamefont{Korotin}}, \bibinfo{journal}{JETP Lett.}
  \textbf{\bibinfo{volume}{88}}, \bibinfo{pages}{729} (\bibinfo{year}{2008}).

\bibitem[{\citenamefont{Yang et~al.}(2009)\citenamefont{Yang, Sorini, Chen,
  Moritz, Lee, Vernay, Olalde-Velasco, Denlinger, Delley, Chu
  et~al.}}]{Yang+09}
\bibinfo{author}{\bibfnamefont{W.~L.} \bibnamefont{Yang}},
  \bibinfo{author}{\bibfnamefont{A.~P.} \bibnamefont{Sorini}},
  \bibinfo{author}{\bibfnamefont{C.-C.} \bibnamefont{Chen}},
  \bibinfo{author}{\bibfnamefont{B.}~\bibnamefont{Moritz}},
  \bibinfo{author}{\bibfnamefont{W.-S.} \bibnamefont{Lee}},
  \bibinfo{author}{\bibfnamefont{F.}~\bibnamefont{Vernay}},
  \bibinfo{author}{\bibfnamefont{P.}~\bibnamefont{Olalde-Velasco}},
  \bibinfo{author}{\bibfnamefont{J.~D.} \bibnamefont{Denlinger}},
  \bibinfo{author}{\bibfnamefont{B.}~\bibnamefont{Delley}},
  \bibinfo{author}{\bibfnamefont{J.-H.} \bibnamefont{Chu}},
  \bibnamefont{et~al.}, \bibinfo{journal}{Phys. Rev. B}
  \textbf{\bibinfo{volume}{80}}, \bibinfo{pages}{014508}
  (\bibinfo{year}{2009}).

\bibitem[{\citenamefont{Si and Abrahams}(2008)}]{Si+08}
\bibinfo{author}{\bibfnamefont{Q.}~\bibnamefont{Si}} \bibnamefont{and}
  \bibinfo{author}{\bibfnamefont{E.}~\bibnamefont{Abrahams}},
  \bibinfo{journal}{Phys. Rev. Lett.} \textbf{\bibinfo{volume}{101}},
  \bibinfo{pages}{076401} (\bibinfo{year}{2008}).

\bibitem[{\citenamefont{Fang et~al.}(2008)\citenamefont{Fang, Yao, Tsai, Hu,
  and Kivelson}}]{Fang+08}
\bibinfo{author}{\bibfnamefont{C.}~\bibnamefont{Fang}},
  \bibinfo{author}{\bibfnamefont{H.}~\bibnamefont{Yao}},
  \bibinfo{author}{\bibfnamefont{W.-F.} \bibnamefont{Tsai}},
  \bibinfo{author}{\bibfnamefont{J.~P.} \bibnamefont{Hu}}, \bibnamefont{and}
  \bibinfo{author}{\bibfnamefont{S.~A.} \bibnamefont{Kivelson}},
  \bibinfo{journal}{Phys. Rev. B} \textbf{\bibinfo{volume}{77}},
  \bibinfo{pages}{224509} (\bibinfo{year}{2008}).

\bibitem[{\citenamefont{Xu et~al.}(2008)\citenamefont{Xu, Mueller, and
  Sachdev}}]{Xu+08b}
\bibinfo{author}{\bibfnamefont{C.}~\bibnamefont{Xu}},
  \bibinfo{author}{\bibfnamefont{M.}~\bibnamefont{Mueller}}, \bibnamefont{and}
  \bibinfo{author}{\bibfnamefont{S.}~\bibnamefont{Sachdev}},
  \bibinfo{journal}{Phys. Rev. B} \textbf{\bibinfo{volume}{78}},
  \bibinfo{pages}{020501(R)} (\bibinfo{year}{2008}).

\bibitem[{\citenamefont{Mazin et~al.}(2008)\citenamefont{Mazin, Singh,
  Johannes, and Du}}]{Mazin+08}
\bibinfo{author}{\bibfnamefont{I.~I.} \bibnamefont{Mazin}},
  \bibinfo{author}{\bibfnamefont{D.~J.} \bibnamefont{Singh}},
  \bibinfo{author}{\bibfnamefont{M.~D.} \bibnamefont{Johannes}},
  \bibnamefont{and} \bibinfo{author}{\bibfnamefont{M.~H.} \bibnamefont{Du}},
  \bibinfo{journal}{Phys. Rev. Lett.} \textbf{\bibinfo{volume}{101}},
  \bibinfo{pages}{057003} (\bibinfo{year}{2008}).

\bibitem[{\citenamefont{Kuroki et~al.}(2008)\citenamefont{Kuroki, Onari, Arita,
  Usui, Tanaka, Kontani, and Aoki}}]{Kuroki+08}
\bibinfo{author}{\bibfnamefont{K.}~\bibnamefont{Kuroki}},
  \bibinfo{author}{\bibfnamefont{S.}~\bibnamefont{Onari}},
  \bibinfo{author}{\bibfnamefont{R.}~\bibnamefont{Arita}},
  \bibinfo{author}{\bibfnamefont{H.}~\bibnamefont{Usui}},
  \bibinfo{author}{\bibfnamefont{Y.}~\bibnamefont{Tanaka}},
  \bibinfo{author}{\bibfnamefont{H.}~\bibnamefont{Kontani}}, \bibnamefont{and}
  \bibinfo{author}{\bibfnamefont{H.}~\bibnamefont{Aoki}},
  \bibinfo{journal}{Phys. Rev. Lett.} \textbf{\bibinfo{volume}{101}},
  \bibinfo{pages}{087004} (\bibinfo{year}{2008}).

\bibitem[{\citenamefont{Raghu et~al.}(2008)\citenamefont{Raghu, Qi, Liu,
  Scalapino, and Zhang}}]{Raghu+08}
\bibinfo{author}{\bibfnamefont{S.}~\bibnamefont{Raghu}},
  \bibinfo{author}{\bibfnamefont{X.-L.} \bibnamefont{Qi}},
  \bibinfo{author}{\bibfnamefont{C.-X.} \bibnamefont{Liu}},
  \bibinfo{author}{\bibfnamefont{D.~J.} \bibnamefont{Scalapino}},
  \bibnamefont{and} \bibinfo{author}{\bibfnamefont{S.-C.} \bibnamefont{Zhang}},
  \bibinfo{journal}{Phys. Rev. B} \textbf{\bibinfo{volume}{77}},
  \bibinfo{pages}{220503(R)} (\bibinfo{year}{2008}).

\bibitem[{\citenamefont{Chubukov et~al.}(2008)\citenamefont{Chubukov, Efremov,
  and Eremin}}]{Chubukov+08}
\bibinfo{author}{\bibfnamefont{A.~V.} \bibnamefont{Chubukov}},
  \bibinfo{author}{\bibfnamefont{D.~V.} \bibnamefont{Efremov}},
  \bibnamefont{and} \bibinfo{author}{\bibfnamefont{I.}~\bibnamefont{Eremin}},
  \bibinfo{journal}{Phys. Rev. B} \textbf{\bibinfo{volume}{78}},
  \bibinfo{pages}{134512} (\bibinfo{year}{2008}).

\bibitem[{\citenamefont{Cvetkovic and Tesanovic}(2009)}]{Cvetkovic+09}
\bibinfo{author}{\bibfnamefont{V.}~\bibnamefont{Cvetkovic}} \bibnamefont{and}
  \bibinfo{author}{\bibfnamefont{Z.}~\bibnamefont{Tesanovic}},
  \bibinfo{journal}{Europhys. Lett.} \textbf{\bibinfo{volume}{85}},
  \bibinfo{pages}{37002} (\bibinfo{year}{2009}).

\bibitem[{\citenamefont{Wang et~al.}(2009)\citenamefont{Wang, Zhai, Ran,
  Vishwanath, and Lee}}]{Wang+09}
\bibinfo{author}{\bibfnamefont{F.}~\bibnamefont{Wang}},
  \bibinfo{author}{\bibfnamefont{H.}~\bibnamefont{Zhai}},
  \bibinfo{author}{\bibfnamefont{Y.}~\bibnamefont{Ran}},
  \bibinfo{author}{\bibfnamefont{A.}~\bibnamefont{Vishwanath}},
  \bibnamefont{and} \bibinfo{author}{\bibfnamefont{D.-H.} \bibnamefont{Lee}},
  \bibinfo{journal}{Phys. Rev. Lett.} \textbf{\bibinfo{volume}{102}},
  \bibinfo{pages}{047005} (\bibinfo{year}{2009}).

\bibitem[{\citenamefont{Brydon and Timm}(2009)}]{Brydon+09}
\bibinfo{author}{\bibfnamefont{P.~M.~R.} \bibnamefont{Brydon}}
  \bibnamefont{and} \bibinfo{author}{\bibfnamefont{C.}~\bibnamefont{Timm}},
  \bibinfo{journal}{Phys. Rev. B} \textbf{\bibinfo{volume}{80}},
  \bibinfo{pages}{174401} (\bibinfo{year}{2009}).

\bibitem[{\citenamefont{Knolle et~al.}(2010)\citenamefont{Knolle, Eremin,
  Chubukov, and Moessner}}]{Knolle+10}
\bibinfo{author}{\bibfnamefont{J.}~\bibnamefont{Knolle}},
  \bibinfo{author}{\bibfnamefont{I.}~\bibnamefont{Eremin}},
  \bibinfo{author}{\bibfnamefont{A.~V.} \bibnamefont{Chubukov}},
  \bibnamefont{and} \bibinfo{author}{\bibfnamefont{R.}~\bibnamefont{Moessner}},
  \bibinfo{journal}{Phys. Rev. B} \textbf{\bibinfo{volume}{81}},
  \bibinfo{pages}{140506} (\bibinfo{year}{2010}).

\bibitem[{\citenamefont{Zhang et~al.}(2010)\citenamefont{Zhang, Opahle,
  Jeschke, and Valenti}}]{Zhang+10}
\bibinfo{author}{\bibfnamefont{Y.-Z.} \bibnamefont{Zhang}},
  \bibinfo{author}{\bibfnamefont{I.}~\bibnamefont{Opahle}},
  \bibinfo{author}{\bibfnamefont{H.~O.} \bibnamefont{Jeschke}},
  \bibnamefont{and} \bibinfo{author}{\bibfnamefont{R.}~\bibnamefont{Valenti}},
  \bibinfo{journal}{Phys. Rev. B} \textbf{\bibinfo{volume}{81}},
  \bibinfo{pages}{094505} (\bibinfo{year}{2010}).

\bibitem[{\citenamefont{Yildirim}(2008)}]{Yildirim+08}
\bibinfo{author}{\bibfnamefont{T.}~\bibnamefont{Yildirim}},
  \bibinfo{journal}{Phys. Rev. Lett.} \textbf{\bibinfo{volume}{101}},
  \bibinfo{pages}{057010} (\bibinfo{year}{2008}).

\bibitem[{\citenamefont{Kr\"uger et~al.}(2009)\citenamefont{Kr\"uger, Kumar,
  Zaanen, and van~den Brink}}]{Kruger+09}
\bibinfo{author}{\bibfnamefont{F.}~\bibnamefont{Kr\"uger}},
  \bibinfo{author}{\bibfnamefont{S.}~\bibnamefont{Kumar}},
  \bibinfo{author}{\bibfnamefont{J.}~\bibnamefont{Zaanen}}, \bibnamefont{and}
  \bibinfo{author}{\bibfnamefont{J.}~\bibnamefont{van~den Brink}},
  \bibinfo{journal}{Phys. Rev. B} \textbf{\bibinfo{volume}{79}},
  \bibinfo{pages}{054504} (\bibinfo{year}{2009}).

\bibitem[{\citenamefont{Jiang et~al.}(2009)\citenamefont{Jiang, Kr\"uger,
  Moore, Sheng, Zaanen, and Weng}}]{Jiang+09}
\bibinfo{author}{\bibfnamefont{H.~C.} \bibnamefont{Jiang}},
  \bibinfo{author}{\bibfnamefont{F.}~\bibnamefont{Kr\"uger}},
  \bibinfo{author}{\bibfnamefont{J.~E.} \bibnamefont{Moore}},
  \bibinfo{author}{\bibfnamefont{D.~N.} \bibnamefont{Sheng}},
  \bibinfo{author}{\bibfnamefont{J.}~\bibnamefont{Zaanen}}, \bibnamefont{and}
  \bibinfo{author}{\bibfnamefont{Z.~Y.} \bibnamefont{Weng}},
  \bibinfo{journal}{Phys. Rev. B} \textbf{\bibinfo{volume}{79}},
  \bibinfo{pages}{174409} (\bibinfo{year}{2009}).

\bibitem[{\citenamefont{Uhrig et~al.}(2009)\citenamefont{Uhrig, Holt, Oitmaa,
  Sushkov, and Singh}}]{Uhrig+09}
\bibinfo{author}{\bibfnamefont{G.~S.} \bibnamefont{Uhrig}},
  \bibinfo{author}{\bibfnamefont{M.}~\bibnamefont{Holt}},
  \bibinfo{author}{\bibfnamefont{J.}~\bibnamefont{Oitmaa}},
  \bibinfo{author}{\bibfnamefont{O.~P.} \bibnamefont{Sushkov}},
  \bibnamefont{and} \bibinfo{author}{\bibfnamefont{R.~R.~P.}
  \bibnamefont{Singh}}, \bibinfo{journal}{Phys. Rev. B}
  \textbf{\bibinfo{volume}{79}}, \bibinfo{pages}{092416}
  (\bibinfo{year}{2009}).

\bibitem[{\citenamefont{Applegate et~al.}(2010)\citenamefont{Applegate, Oitmaa,
  and Singh}}]{Applegate+10}
\bibinfo{author}{\bibfnamefont{R.}~\bibnamefont{Applegate}},
  \bibinfo{author}{\bibfnamefont{J.}~\bibnamefont{Oitmaa}}, \bibnamefont{and}
  \bibinfo{author}{\bibfnamefont{R.~R.~P.} \bibnamefont{Singh}},
  \bibinfo{journal}{Phys. Rev. B} \textbf{\bibinfo{volume}{81}},
  \bibinfo{pages}{024505} (\bibinfo{year}{2010}).

\bibitem[{\citenamefont{Schmidt et~al.}(2010)\citenamefont{Schmidt, Siahatgar,
  and Thalmeier}}]{Schmidt+10}
\bibinfo{author}{\bibfnamefont{B.}~\bibnamefont{Schmidt}},
  \bibinfo{author}{\bibfnamefont{M.}~\bibnamefont{Siahatgar}},
  \bibnamefont{and}
  \bibinfo{author}{\bibfnamefont{P.}~\bibnamefont{Thalmeier}},
  \bibinfo{journal}{Phys. Rev. B} \textbf{\bibinfo{volume}{81}},
  \bibinfo{pages}{165101} (\bibinfo{year}{2010}).

\bibitem[{\citenamefont{Zhao et~al.}(2008{\natexlab{d}})\citenamefont{Zhao,
  Yao, Li, Hong, Chen, Chang, II, Lynn, Mook, Chen et~al.}}]{Zhao+08}
\bibinfo{author}{\bibfnamefont{J.}~\bibnamefont{Zhao}},
  \bibinfo{author}{\bibfnamefont{D.-X.} \bibnamefont{Yao}},
  \bibinfo{author}{\bibfnamefont{S.}~\bibnamefont{Li}},
  \bibinfo{author}{\bibfnamefont{T.}~\bibnamefont{Hong}},
  \bibinfo{author}{\bibfnamefont{Y.}~\bibnamefont{Chen}},
  \bibinfo{author}{\bibfnamefont{S.}~\bibnamefont{Chang}},
  \bibinfo{author}{\bibfnamefont{W.~R.} \bibnamefont{II}},
  \bibinfo{author}{\bibfnamefont{J.~W.} \bibnamefont{Lynn}},
  \bibinfo{author}{\bibfnamefont{H.~A.} \bibnamefont{Mook}},
  \bibinfo{author}{\bibfnamefont{G.~F.} \bibnamefont{Chen}},
  \bibnamefont{et~al.}, \bibinfo{journal}{Phys. Rev. Lett.}
  \textbf{\bibinfo{volume}{101}}, \bibinfo{pages}{167203}
  (\bibinfo{year}{2008}{\natexlab{d}}).

\bibitem[{\citenamefont{Zhao et~al.}(2009)\citenamefont{Zhao, Adroja, Yao,
  Bewley, Li, Wang, Wu, Chen, Hu, and Dai}}]{Zhao+09}
\bibinfo{author}{\bibfnamefont{J.}~\bibnamefont{Zhao}},
  \bibinfo{author}{\bibfnamefont{D.~T.} \bibnamefont{Adroja}},
  \bibinfo{author}{\bibfnamefont{D.-X.} \bibnamefont{Yao}},
  \bibinfo{author}{\bibfnamefont{R.}~\bibnamefont{Bewley}},
  \bibinfo{author}{\bibfnamefont{S.}~\bibnamefont{Li}},
  \bibinfo{author}{\bibfnamefont{X.~F.} \bibnamefont{Wang}},
  \bibinfo{author}{\bibfnamefont{G.}~\bibnamefont{Wu}},
  \bibinfo{author}{\bibfnamefont{X.~H.} \bibnamefont{Chen}},
  \bibinfo{author}{\bibfnamefont{J.}~\bibnamefont{Hu}}, \bibnamefont{and}
  \bibinfo{author}{\bibfnamefont{P.}~\bibnamefont{Dai}},
  \bibinfo{journal}{Nature Physics} \textbf{\bibinfo{volume}{5}},
  \bibinfo{pages}{555} (\bibinfo{year}{2009}).

\bibitem[{\citenamefont{Chandra and Doucot}(1988)}]{Chandra+88}
\bibinfo{author}{\bibfnamefont{P.}~\bibnamefont{Chandra}} \bibnamefont{and}
  \bibinfo{author}{\bibfnamefont{B.}~\bibnamefont{Doucot}},
  \bibinfo{journal}{Phys. Rev. B} \textbf{\bibinfo{volume}{38}},
  \bibinfo{pages}{9335} (\bibinfo{year}{1988}).

\bibitem[{\citenamefont{Henley}(1989)}]{Henley+89}
\bibinfo{author}{\bibfnamefont{C.~L.} \bibnamefont{Henley}},
  \bibinfo{journal}{Phys. Rev. Lett.} \textbf{\bibinfo{volume}{62}},
  \bibinfo{pages}{2056} (\bibinfo{year}{1989}).

\bibitem[{\citenamefont{Chandra et~al.}(1990)\citenamefont{Chandra, Coleman,
  and Larkin}}]{Chandra+90}
\bibinfo{author}{\bibfnamefont{P.}~\bibnamefont{Chandra}},
  \bibinfo{author}{\bibfnamefont{P.}~\bibnamefont{Coleman}}, \bibnamefont{and}
  \bibinfo{author}{\bibfnamefont{A.~I.} \bibnamefont{Larkin}},
  \bibinfo{journal}{Phys. Rev. Lett.} \textbf{\bibinfo{volume}{64}},
  \bibinfo{pages}{88} (\bibinfo{year}{1990}).

\bibitem[{\citenamefont{Lv et~al.}(2009)\citenamefont{Lv, Wu, and
  Phillips}}]{Lv+09}
\bibinfo{author}{\bibfnamefont{W.}~\bibnamefont{Lv}},
  \bibinfo{author}{\bibfnamefont{J.}~\bibnamefont{Wu}}, \bibnamefont{and}
  \bibinfo{author}{\bibfnamefont{P.}~\bibnamefont{Phillips}},
  \bibinfo{journal}{Phys. Rev. B} \textbf{\bibinfo{volume}{80}},
  \bibinfo{pages}{224506} (\bibinfo{year}{2009}).

\bibitem[{\citenamefont{Lee et~al.}(2009)\citenamefont{Lee, Yin, and
  Ku}}]{Lee+09}
\bibinfo{author}{\bibfnamefont{C.-C.} \bibnamefont{Lee}},
  \bibinfo{author}{\bibfnamefont{W.-G.} \bibnamefont{Yin}}, \bibnamefont{and}
  \bibinfo{author}{\bibfnamefont{W.}~\bibnamefont{Ku}}, \bibinfo{journal}{Phys.
  Rev. Lett} \textbf{\bibinfo{volume}{103}}, \bibinfo{pages}{267001}
  (\bibinfo{year}{2009}).

\bibitem[{\citenamefont{Chen et~al.}(2009)\citenamefont{Chen, Moritz, van~den
  Brink, Devereaux, and Singh}}]{Chen+09}
\bibinfo{author}{\bibfnamefont{C.-C.} \bibnamefont{Chen}},
  \bibinfo{author}{\bibfnamefont{B.}~\bibnamefont{Moritz}},
  \bibinfo{author}{\bibfnamefont{J.}~\bibnamefont{van~den Brink}},
  \bibinfo{author}{\bibfnamefont{T.~P.} \bibnamefont{Devereaux}},
  \bibnamefont{and} \bibinfo{author}{\bibfnamefont{R.~R.~P.}
  \bibnamefont{Singh}}, \bibinfo{journal}{Phys. Rev. B}
  \textbf{\bibinfo{volume}{80}}, \bibinfo{pages}{180418}
  (\bibinfo{year}{2009}).

\bibitem[{\citenamefont{Lee and Wu}(2009)}]{Lee+09b}
\bibinfo{author}{\bibfnamefont{W.-C.} \bibnamefont{Lee}} \bibnamefont{and}
  \bibinfo{author}{\bibfnamefont{C.}~\bibnamefont{Wu}}, \bibinfo{journal}{Phys.
  Rev. Lett.} \textbf{\bibinfo{volume}{103}}, \bibinfo{pages}{176101}
  (\bibinfo{year}{2009}).

\bibitem[{\citenamefont{Chuang et~al.}(2010)\citenamefont{Chuang, Allan, Lee,
  Xie, Ni, BudÕko, Boebinger, Canfield, and Davis}}]{Chuang+10}
\bibinfo{author}{\bibfnamefont{T.-M.} \bibnamefont{Chuang}},
  \bibinfo{author}{\bibfnamefont{M.~P.} \bibnamefont{Allan}},
  \bibinfo{author}{\bibfnamefont{J.}~\bibnamefont{Lee}},
  \bibinfo{author}{\bibfnamefont{Y.}~\bibnamefont{Xie}},
  \bibinfo{author}{\bibfnamefont{N.}~\bibnamefont{Ni}},
  \bibinfo{author}{\bibfnamefont{S.~L.} \bibnamefont{BudÕko}},
  \bibinfo{author}{\bibfnamefont{G.~S.} \bibnamefont{Boebinger}},
  \bibinfo{author}{\bibfnamefont{P.~C.} \bibnamefont{Canfield}},
  \bibnamefont{and} \bibinfo{author}{\bibfnamefont{J.~C.} \bibnamefont{Davis}},
  \bibinfo{journal}{Science} \textbf{\bibinfo{volume}{327}},
  \bibinfo{pages}{181} (\bibinfo{year}{2010}).

\bibitem[{\citenamefont{Tanatar et~al.}(2010)\citenamefont{Tanatar, Blomberg,
  Kreyssig, Kim, Ni, Thaler, BudÕko, Canfield, Goldman, Mazin
  et~al.}}]{Tanatar+10}
\bibinfo{author}{\bibfnamefont{M.~A.} \bibnamefont{Tanatar}},
  \bibinfo{author}{\bibfnamefont{E.~C.} \bibnamefont{Blomberg}},
  \bibinfo{author}{\bibfnamefont{A.}~\bibnamefont{Kreyssig}},
  \bibinfo{author}{\bibfnamefont{M.~G.} \bibnamefont{Kim}},
  \bibinfo{author}{\bibfnamefont{N.}~\bibnamefont{Ni}},
  \bibinfo{author}{\bibfnamefont{A.}~\bibnamefont{Thaler}},
  \bibinfo{author}{\bibfnamefont{S.~L.} \bibnamefont{BudÕko}},
  \bibinfo{author}{\bibfnamefont{P.~C.} \bibnamefont{Canfield}},
  \bibinfo{author}{\bibfnamefont{A.~I.} \bibnamefont{Goldman}},
  \bibinfo{author}{\bibfnamefont{I.~I.} \bibnamefont{Mazin}},
  \bibnamefont{et~al.}, \bibinfo{journal}{Phys. Rev. B}
  \textbf{\bibinfo{volume}{81}}, \bibinfo{pages}{184508}
  (\bibinfo{year}{2010}).

\bibitem[{\citenamefont{Chu et~al.}(2010)\citenamefont{Chu, Analytis, Greve,
  McMahon, Islam, Yamamoto, and Fisher}}]{Chu+10}
\bibinfo{author}{\bibfnamefont{J.-H.} \bibnamefont{Chu}},
  \bibinfo{author}{\bibfnamefont{J.~G.} \bibnamefont{Analytis}},
  \bibinfo{author}{\bibfnamefont{K.~D.} \bibnamefont{Greve}},
  \bibinfo{author}{\bibfnamefont{P.~L.} \bibnamefont{McMahon}},
  \bibinfo{author}{\bibfnamefont{Z.}~\bibnamefont{Islam}},
  \bibinfo{author}{\bibfnamefont{Y.}~\bibnamefont{Yamamoto}}, \bibnamefont{and}
  \bibinfo{author}{\bibfnamefont{I.~R.} \bibnamefont{Fisher}}
  (\bibinfo{year}{2010}), \eprint{arXiv:1002.3364}.

\bibitem[{\citenamefont{Shimojima et~al.}(2010)\citenamefont{Shimojima,
  Ishizaka, Ishida, Katayama, Ohgushi, Kiss, Okawa, Togashi, Wang, Chen
  et~al.}}]{Shimojima+10}
\bibinfo{author}{\bibfnamefont{T.}~\bibnamefont{Shimojima}},
  \bibinfo{author}{\bibfnamefont{K.}~\bibnamefont{Ishizaka}},
  \bibinfo{author}{\bibfnamefont{Y.}~\bibnamefont{Ishida}},
  \bibinfo{author}{\bibfnamefont{N.}~\bibnamefont{Katayama}},
  \bibinfo{author}{\bibfnamefont{K.}~\bibnamefont{Ohgushi}},
  \bibinfo{author}{\bibfnamefont{T.}~\bibnamefont{Kiss}},
  \bibinfo{author}{\bibfnamefont{M.}~\bibnamefont{Okawa}},
  \bibinfo{author}{\bibfnamefont{T.}~\bibnamefont{Togashi}},
  \bibinfo{author}{\bibfnamefont{X.-Y.} \bibnamefont{Wang}},
  \bibinfo{author}{\bibfnamefont{C.-T.} \bibnamefont{Chen}},
  \bibnamefont{et~al.}, \bibinfo{journal}{Phys. Rev. Lett}
  \textbf{\bibinfo{volume}{104}}, \bibinfo{pages}{057002}
  (\bibinfo{year}{2010}).

\bibitem[{\citenamefont{McGuire et~al.}(2008)\citenamefont{McGuire,
  Christianson, Sefat, Sales, Lumsden, Jin, Payzant, Mandrus, Luan, Keppens
  et~al.}}]{McGuire+08}
\bibinfo{author}{\bibfnamefont{M.~A.} \bibnamefont{McGuire}},
  \bibinfo{author}{\bibfnamefont{A.~D.} \bibnamefont{Christianson}},
  \bibinfo{author}{\bibfnamefont{A.~S.} \bibnamefont{Sefat}},
  \bibinfo{author}{\bibfnamefont{B.~C.} \bibnamefont{Sales}},
  \bibinfo{author}{\bibfnamefont{M.~D.} \bibnamefont{Lumsden}},
  \bibinfo{author}{\bibfnamefont{R.}~\bibnamefont{Jin}},
  \bibinfo{author}{\bibfnamefont{E.~A.} \bibnamefont{Payzant}},
  \bibinfo{author}{\bibfnamefont{D.}~\bibnamefont{Mandrus}},
  \bibinfo{author}{\bibfnamefont{Y.}~\bibnamefont{Luan}},
  \bibinfo{author}{\bibfnamefont{V.}~\bibnamefont{Keppens}},
  \bibnamefont{et~al.}, \bibinfo{journal}{Phys. Rev. B}
  \textbf{\bibinfo{volume}{78}}, \bibinfo{pages}{094517}
  (\bibinfo{year}{2008}).

\bibitem[{\citenamefont{Akrap et~al.}(2009)\citenamefont{Akrap, Tu, Li, Cao,
  Xu, and Homes}}]{Akrap+09}
\bibinfo{author}{\bibfnamefont{A.}~\bibnamefont{Akrap}},
  \bibinfo{author}{\bibfnamefont{J.~J.} \bibnamefont{Tu}},
  \bibinfo{author}{\bibfnamefont{L.~J.} \bibnamefont{Li}},
  \bibinfo{author}{\bibfnamefont{G.~H.} \bibnamefont{Cao}},
  \bibinfo{author}{\bibfnamefont{Z.~A.} \bibnamefont{Xu}}, \bibnamefont{and}
  \bibinfo{author}{\bibfnamefont{C.~C.} \bibnamefont{Homes}},
  \bibinfo{journal}{Phys. Rev. B} \textbf{\bibinfo{volume}{80}},
  \bibinfo{pages}{180502(R)} (\bibinfo{year}{2009}).

\bibitem[{\citenamefont{Kou et~al.}(2009)\citenamefont{Kou, Li, and
  Weng}}]{Kou+09}
\bibinfo{author}{\bibfnamefont{S.-P.} \bibnamefont{Kou}},
  \bibinfo{author}{\bibfnamefont{T.}~\bibnamefont{Li}}, \bibnamefont{and}
  \bibinfo{author}{\bibfnamefont{Z.-Y.} \bibnamefont{Weng}},
  \bibinfo{journal}{Europhys. Lett.} \textbf{\bibinfo{volume}{88}},
  \bibinfo{pages}{17010} (\bibinfo{year}{2009}).

\bibitem[{\citenamefont{Dai et~al.}(2009)\citenamefont{Dai, Si, Zhu, and
  Abrahams}}]{Dai+09}
\bibinfo{author}{\bibfnamefont{J.}~\bibnamefont{Dai}},
  \bibinfo{author}{\bibfnamefont{Q.}~\bibnamefont{Si}},
  \bibinfo{author}{\bibfnamefont{J.-X.} \bibnamefont{Zhu}}, \bibnamefont{and}
  \bibinfo{author}{\bibfnamefont{E.}~\bibnamefont{Abrahams}},
  \bibinfo{journal}{Proc. Nat. Acad. Sci.} \textbf{\bibinfo{volume}{106}},
  \bibinfo{pages}{4118} (\bibinfo{year}{2009}).

\bibitem[{\citenamefont{van~den Brink and Khomskii}(1999)}]{Brink+99a}
\bibinfo{author}{\bibfnamefont{J.}~\bibnamefont{van~den Brink}}
  \bibnamefont{and} \bibinfo{author}{\bibfnamefont{D.}~\bibnamefont{Khomskii}},
  \bibinfo{journal}{Phys. Rev. Lett.} \textbf{\bibinfo{volume}{82}},
  \bibinfo{pages}{1016} (\bibinfo{year}{1999}).

\bibitem[{\citenamefont{van~den Brink et~al.}(1999)\citenamefont{van~den Brink,
  Khaliullin, and Khomskii}}]{Brink+99b}
\bibinfo{author}{\bibfnamefont{J.}~\bibnamefont{van~den Brink}},
  \bibinfo{author}{\bibfnamefont{G.}~\bibnamefont{Khaliullin}},
  \bibnamefont{and} \bibinfo{author}{\bibfnamefont{D.}~\bibnamefont{Khomskii}},
  \bibinfo{journal}{Phys. Rev. Lett.} \textbf{\bibinfo{volume}{83}},
  \bibinfo{pages}{5118} (\bibinfo{year}{1999}).

\bibitem[{\citenamefont{van~den Brink and Khomskii}(2001)}]{Brink+01}
\bibinfo{author}{\bibfnamefont{J.}~\bibnamefont{van~den Brink}}
  \bibnamefont{and} \bibinfo{author}{\bibfnamefont{D.}~\bibnamefont{Khomskii}},
  \bibinfo{journal}{Phys. Rev. B} \textbf{\bibinfo{volume}{63}},
  \bibinfo{pages}{140416(R)} (\bibinfo{year}{2001}).

\bibitem[{\citenamefont{Senff et~al.}(2006)\citenamefont{Senff, Kr\"uger,
  Scheidl, Benomar, Sidis, Demmel, and Braden}}]{Senff+06}
\bibinfo{author}{\bibfnamefont{D.}~\bibnamefont{Senff}},
  \bibinfo{author}{\bibfnamefont{F.}~\bibnamefont{Kr\"uger}},
  \bibinfo{author}{\bibfnamefont{S.}~\bibnamefont{Scheidl}},
  \bibinfo{author}{\bibfnamefont{M.}~\bibnamefont{Benomar}},
  \bibinfo{author}{\bibfnamefont{Y.}~\bibnamefont{Sidis}},
  \bibinfo{author}{\bibfnamefont{F.}~\bibnamefont{Demmel}}, \bibnamefont{and}
  \bibinfo{author}{\bibfnamefont{M.}~\bibnamefont{Braden}},
  \bibinfo{journal}{Phys. Rev. Lett.} \textbf{\bibinfo{volume}{96}},
  \bibinfo{pages}{257201} (\bibinfo{year}{2006}).

\bibitem[{\citenamefont{Neuber et~al.}(2006)\citenamefont{Neuber, Daghofer,
  Ole\'{s}, and von~der Linden}}]{Neuber+06}
\bibinfo{author}{\bibfnamefont{D.~R.} \bibnamefont{Neuber}},
  \bibinfo{author}{\bibfnamefont{M.}~\bibnamefont{Daghofer}},
  \bibinfo{author}{\bibfnamefont{A.~M.} \bibnamefont{Ole\'{s}}},
  \bibnamefont{and} \bibinfo{author}{\bibfnamefont{W.}~\bibnamefont{von~der
  Linden}}, \bibinfo{journal}{Phys. Stat. Sol. (c)}
  \textbf{\bibinfo{volume}{3}}, \bibinfo{pages}{32} (\bibinfo{year}{2006}).

\bibitem[{\citenamefont{Ole\'{s}}(1983)}]{Oles+83}
\bibinfo{author}{\bibfnamefont{A.~M.} \bibnamefont{Ole\'{s}}},
  \bibinfo{journal}{Phys. Rev. B} \textbf{\bibinfo{volume}{28}},
  \bibinfo{pages}{327} (\bibinfo{year}{1983}).

\bibitem[{\citenamefont{Moreo et~al.}(2009)\citenamefont{Moreo, Daghofer,
  Riera, and Dagotto}}]{Moreo+09}
\bibinfo{author}{\bibfnamefont{A.}~\bibnamefont{Moreo}},
  \bibinfo{author}{\bibfnamefont{M.}~\bibnamefont{Daghofer}},
  \bibinfo{author}{\bibfnamefont{J.~A.} \bibnamefont{Riera}}, \bibnamefont{and}
  \bibinfo{author}{\bibfnamefont{E.}~\bibnamefont{Dagotto}},
  \bibinfo{journal}{Phys. Rev. B} \textbf{\bibinfo{volume}{79}},
  \bibinfo{pages}{134502} (\bibinfo{year}{2009}).

\bibitem[{\citenamefont{Nagaev}(1998)}]{Nagaev+98}
\bibinfo{author}{\bibfnamefont{E.~L.} \bibnamefont{Nagaev}},
  \bibinfo{journal}{Phys. Rev. B} \textbf{\bibinfo{volume}{58}},
  \bibinfo{pages}{827} (\bibinfo{year}{1998}).

\bibitem[{\citenamefont{Golosov}(2000)}]{Golosov+00}
\bibinfo{author}{\bibfnamefont{D.~I.} \bibnamefont{Golosov}},
  \bibinfo{journal}{Phys. Rev. Lett.} \textbf{\bibinfo{volume}{84}},
  \bibinfo{pages}{3974} (\bibinfo{year}{2000}).

\bibitem[{\citenamefont{Shannon and Chubukov}(2002)}]{Shannon+02}
\bibinfo{author}{\bibfnamefont{N.}~\bibnamefont{Shannon}} \bibnamefont{and}
  \bibinfo{author}{\bibfnamefont{A.~V.} \bibnamefont{Chubukov}},
  \bibinfo{journal}{Phys. Rev. B} \textbf{\bibinfo{volume}{65}},
  \bibinfo{pages}{104418} (\bibinfo{year}{2002}).

\bibitem[{\citenamefont{Bascones et~al.}(2010)\citenamefont{Bascones,
  Calder\'on, and Valenzuela}}]{Bascones+10}
\bibinfo{author}{\bibfnamefont{E.}~\bibnamefont{Bascones}},
  \bibinfo{author}{\bibfnamefont{M.~J.} \bibnamefont{Calder\'on}},
  \bibnamefont{and}
  \bibinfo{author}{\bibfnamefont{B.}~\bibnamefont{Valenzuela}},
  \bibinfo{journal}{Phys. Rev. Lett.} \textbf{\bibinfo{volume}{104}},
  \bibinfo{pages}{227201} (\bibinfo{year}{2010}).

\bibitem[{\citenamefont{Daghofer et~al.}(2010)\citenamefont{Daghofer, Luo, Yu,
  Yao, Moreo, and Dagotto}}]{Daghofer+10}
\bibinfo{author}{\bibfnamefont{M.}~\bibnamefont{Daghofer}},
  \bibinfo{author}{\bibfnamefont{Q.-L.} \bibnamefont{Luo}},
  \bibinfo{author}{\bibfnamefont{R.}~\bibnamefont{Yu}},
  \bibinfo{author}{\bibfnamefont{D.~X.} \bibnamefont{Yao}},
  \bibinfo{author}{\bibfnamefont{A.}~\bibnamefont{Moreo}}, \bibnamefont{and}
  \bibinfo{author}{\bibfnamefont{E.}~\bibnamefont{Dagotto}},
  \bibinfo{journal}{Phys. Rev. B} \textbf{\bibinfo{volume}{81}},
  \bibinfo{pages}{180514} (\bibinfo{year}{2010}).

\bibitem[{\citenamefont{Chen et~al.}(2010)\citenamefont{Chen, Maciejko, Sorini,
  Moritz, Singh, and Devereaux}}]{Chen+10}
\bibinfo{author}{\bibfnamefont{C.~C.} \bibnamefont{Chen}},
  \bibinfo{author}{\bibfnamefont{J.}~\bibnamefont{Maciejko}},
  \bibinfo{author}{\bibfnamefont{A.~P.} \bibnamefont{Sorini}},
  \bibinfo{author}{\bibfnamefont{B.}~\bibnamefont{Moritz}},
  \bibinfo{author}{\bibfnamefont{R.~R.~P.} \bibnamefont{Singh}},
  \bibnamefont{and} \bibinfo{author}{\bibfnamefont{T.~P.}
  \bibnamefont{Devereaux}} (\bibinfo{year}{2010}), \eprint{arXiv:1004.4611}.

\end{thebibliography}
\end{document}